\documentclass[a4paper,fleqn,usenatbib]{mnras}

\usepackage[T1]{fontenc}
\usepackage{ae,aecompl}
\usepackage{soul}

\usepackage{graphicx}	
\usepackage{amsmath}	
\usepackage{amssymb}	
\usepackage{float}
\usepackage{dblfloatfix}
\usepackage{subfigure}

\newcommand{\Ha}{H$\alpha$}
\newcommand{\Hb}{H$\beta$}

\newcommand{\Ms}{$M_{\star}$}

\usepackage[dvipsnames]{xcolor}
\usepackage[T1]{fontenc}
\usepackage{ae,aecompl}

\title[Scaling relations in GAMA groups]{Galaxy And Mass Assembly (GAMA):  The environmental impact on SFR and metallicity in galaxy groups}

\author[D. Sotillo-Ramos et al.]{
D. Sotillo-Ramos,$^{1}$\thanks{E-mail: sotillo@mpia.de}
M. A. Lara-L\'opez,$^{2,3}$
A.M. P\'erez-Garc\'ia,$^{4,5}$
R. P\'erez-Mart\'inez,$^{5,6}$
\newauthor
A. M. Hopkins,$^{7}$
B. W. Holwerda,$^{8}$
J. Liske,$^{9}$
A. R. L\'opez-S\'anchez$^{7,10,11,12}$
\newauthor
M. S. Owers,$^{10,11}$ 
K. A. Pimbblet,$^{13}$\\
$^{1}$Max-Planck-Institut f\"ur Astronomie, K\"{o}nigstuhl 17
69117 Heidelberg, Germany\\
$^{2}$Armagh Observatory and Planetarium, College Hill, Armagh, BT61 DG, UK\\
$^{3}$DARK, Niels Bohr Institute, University of Copenhagen, Lyngbyvej 2, Copenhagen DK-2100, Denmark\\
$^{4}$ Centro de Astrobiolog\'ia, CAB/INTA-CSIC, 28692 ESAC Campus, Villanueva de la Ca\~nada, Madrid, Spain\\
$^{5}$ Asociaci\'on Astrof\'isica para la Promoci\'on de la Investigaci\'on, Instrumentaci\'on y su Desarrollo, ASPID, 38205 La Laguna, Tenerife, Spain\\
$^{6}$ ISDEFE for ESA. Camino Bajo del Castillo s/n. Urb. Villafranca del Castillo. E-28692, Villanueva de la Ca\~nada, Spain\\
$^{7}$ Australian Astronomical Optics, Macquarie University, 105 Delhi Rd, North Ryde, NSW 2113, Australia\\
$^{8}$Department of Physics and Astronomy, 102 National Science Building, University of Louisville, Louisville KY 40292, USA\\
$^{9}$Hamburger Sternw{\"a}rte, Universitat Hamburg, Gojenbergsweg 112, 21029 Hamburg, Germany\\
$^{10}$Department of Physics and Astronomy, Macquarie University, NSW 2109, Australia\\
$^{11}$Macquarie University Research Centre for Astronomy, Astrophysics \& Astrophotonics, Sydney, NSW 2109, Australia\\ 
$^{12}$ARC Centre of Excellence for All Sky Astrophysics in 3 Dimensions (ASTRO-3D)\\
$^{13}$E.A.Milne Centre for Astrophysics, University of Hull, Cottingham Road, Kingston-upon-Hull, HU6 7RX, UK\\
}

\date{Accepted XXX. Received YYY; in original form ZZZ}

\pubyear{2019}

\begin{document}
\label{firstpage}
\pagerange{\pageref{firstpage}--\pageref{lastpage}}
\maketitle

\begin{abstract}
We present a study of the relationships and environmental dependencies between stellar mass,
star formation rate, and gas metallicity for more than 700 galaxies in groups up to redshift 0.35 from the Galaxy And Mass Assembly (GAMA) survey. 
To identify the main drivers, our sample was analyzed as a function of group-centric distance, projected galaxy number density, and stellar mass. By using control samples of more than 16000 star-forming field galaxies and volume limited samples, we find that the highest enhancement in SFR (0.3 dex) occurs in galaxies with the lowest local density. 
In contrast to previous work, our data show small enhancements of $\sim$0.1 dex in SFR for galaxies at the highest local densities or group-centric distances.  
Our data indicates quenching in SFR only for massive galaxies, suggesting that stellar mass might be the main driver of quenching processes for star forming galaxies. We can discard a morphological driven quenching, since the S\'ersic index distribution for group and control galaxies are similar.
The gas metallicity does not vary drastically. It increases $\sim$0.08 dex for galaxies at the highest local densities, and decreases for galaxies at the highest group-centric distances, in agreement with previous work.
Altogether, the local density, rather than group-centric distance, shows the stronger impact in enhancing both, the SFR and gas metallicity.
We applied the same methodology to galaxies from the IllustrisTNG simulations, and although we were able to reproduce the general observational trends, the differences between group and control samples only partially agree with the observations

\end{abstract}

\begin{keywords}
galaxies: abundances -- galaxies: fundamental parameters -- galaxies -- star formation
\end{keywords}



\section{Introduction}


The physical processes driving the evolution of galaxies are a complex and important open question in astronomy. The role played by internal versus external processes, commonly known as nature versus nurture scenarios, has been a matter of debate for decades \citep[e.g.,][]{DiMatteo05, Hopkins06}. 

Observationally, there is evidence that between 50-70$\%$ of the galaxy population is in groups \citep[e.g.,][]{Eke05}. This naturally implies that processes taking place in the group environment can have a significant impact on the evolution of the galaxy population as a whole. Groups and clusters of galaxies have long been considered perfect laboratories to study the effect of feedback processes in galaxies. The effect of these processes is likely to manifest through well known scaling relationships of galaxy properties \citep[e.g.,][]{Tremonti2004, Lara13}.

With the advent of large spectroscopic surveys such as the Sloan Digital Sky survey \citep[SDSS,][]{Abazajian09} and Galaxy And Mass Assembly \citep[GAMA,][]{Liske15}, large advancements in our understanding of the environmental processes have been made. For instance, the interplay between the stellar mass (\Ms) and the gas metallicity  ($Z$) in star forming (SF) galaxies is shown to have a very strong correlation, with massive \mbox{galaxies} showing higher metallicities than less massive \mbox{galaxies}, as quantified through the M-Z relation \citep[e.g.,][]{Lequeux79,Tremonti2004}. The M-Z relation has been extensively \mbox{studied} at local redshifts \citep[e.g.,][]{Tremonti2004,Kewley08}, and metallicity has been shown to evolve to lower values up to redshifts of $z \approx$ 0.4 \citep[e.g.,][]{Lara09a,Lara09b, Lara10a, Pilyugin11}. 


Since metallicity is sensitive to metal losses due to \mbox{stellar} winds \citep{Spitoni10,Tremonti2004}, supernovae \citep{Brooks07}, and active galactic nuclei (AGN) feedback \citep{Lara19}, the M-Z relation provides essential insight into galaxy formation and evolution. Furthermore, the environment also plays an important role in the gas metallicity and properties of galaxies. Galaxy interactions and mergers can cause gas inflows, morphological transformations, trigger star formation and even lead to activity in the galactic nucleus \citep[e.g.,][]{Barton00,Lambas03,Nikolic04,Alonso07,Woods2007, Ellison08, davies15, gordon18, Ellison19, Pan19, Shah20}.

Studies of  galaxies in  pairs and clusters have revealed the environmental effects on the M-Z relation. For instance, several authors \citep[e. g., ][]{Kewley06, Ellison08, Scudder12, Scudder15} find that galaxies in close pairs are more metal poor by approximately $\sim$0.1 dex at a given luminosity compared to galaxies with no near companion. On the other hand, \cite{Ellison09} find that galaxies in clusters tend to have higher \mbox{metallicities} by up to $\sim$0.04 dex when compared to a control sample of the same mass, redshift, fibre covering fraction and rest-frame $g-r$ color. This last study emphasises that the metal enhancements are driven by local overdensities, and not just cluster membership. In terms of redshift, in the local universe galaxies in clusters have higher metallicities on \mbox{average} at given stellar mass  \citep[e.g.,][]{Peng14}, but at higher redshifts the influence of environment on \mbox{metallicity} is not clear \citep{Kulas13,Shimakawa15,Valentino15}. More recently, \citet{Wu17} \mbox{examined} the M-Z relation as a function of \mbox{environment} based  on the analysis  of
$\sim$ 40000  galaxies  in  the  SDSS, and  show that the metallicity has a weak dependence on the environment. Indeed, environmental processes that trigger short-lived bursts of star formation may cause a significant, but transient, change in a galaxy's metallicity before it returns to an equilibrium metallicity \citep{Finlator08}.

Moreover, the study of the relationship between stellar mass and star formation rate (\Ms-SFR) or specific star formation rate (\Ms-sSFR) allows us to understand the influence of environment in the evolution of galaxies and the physical processes at the origin of the quenching of star formation. It is well known that the mass and SFR follow a tight relation  for star-forming galaxies --- the main-sequence (MS) relation --- in both the local and high-redshift universe, with this relation shifting to higher SFR at higher $z$ for a given mass 
\citep[e.g.,][]{wuyts11,karim11}. Nevertheless, few works have focused on the influence of environment on the MS, and contradictory results have been found. Some authors \citep{york00,lilly07,wang18} find that star formation activity in groups and clusters is, on average, reduced with respect to field galaxies, at a given stellar mass, while \citet{peng10} argued that these relations do not depend on environment. Recently, \citet{calvi18} support the conclusion of \citet{peng10}, suggesting that morphology drives the relation between mass and SFR, more than environment.

Furthermore, the M-SFR relation has been used in the analysis of quenching processes in galaxies, where red$/$quiescent galaxies are characterized by denser environments \citep[e.g.,][]{Barlog97, Lewis02, Baldry06, Peng12,mcnaught14}. The processes involved in quenching the star formation on the other hand, are still a matter of debate, and can involve a combination of ram-pressure stripping, starvation, harassment and mergers \citep[e.g.,][]{Gunn72, Moore96, Schawinski14, Peng15,Trussler20}.
Alternatively, a scenario of in situ evolution could play a role, where passive early-type galaxies may have evolved early and rapidly within dense environments, while present-day star-forming galaxies in low-density regions have evolved more slowly, and are yet to be quenched \citep[e.g.,][]{Wijesinghe12}.

Low mass \mbox{galaxies}, however, may be characterized by a delayed-then-rapid quenching scenario, initially proposed for satellite \mbox{galaxies} in clusters \citep[e.g.,][]{Wetzel13, Oman16}, and recently confirmed by \citet[][]{Moutard18} up to redshifts $\sim$ 0.6. In this scenario, the star formation is suppressed for $\sim$ 0.4 Gyr, and associated with a further morphological transformation.  On the other hand, \citet{corcho2020} suggest an interpretation of the M-sSFR relation in terms of a single population of galaxies at different ``ageing" stages, in contrast to the bimodal picture of active and passive galaxies separated by quenching processes.


The M-Z, M-SFR and M-sSFR relations have also been studied by means of cosmological simulations. For instance, \citet{Furlong2015} used the EAGLE \citep[Evolution and Assembly of GaLaxies and their Environments,][]{schaller15} simulation to study the M-SFR relation and the evolution of Z and SFR with time. Compared to observations, they find similar trends but a discrepancy for all ranges of mass and redshift.
\citet{Dave2017}, making use of the MUFASA simulation \citep[Galaxy Formation Simulations With Meshless Hydrodynamics,][]{dave16}, find that galaxies with lower metallicities exhibit higher specific star formation rates, for fixed stellar mass. \citet{DeRossi2018} come to similar results with the EAGLE simulation.
\citet{Bahe2017}, in an analysis similar to that presented in this work, used the same simulations to compare metallicities of satellites and field galaxies, for a fixed redshift z=0.1. They find similar discrepancies with the observations and also an excess in the metallicity of satellite compared to field galaxies.

This paper introduces the relationships of mass, \mbox{metallicity}, SFR, and specific SFR for galaxies in groups in the GAMA survey. In \S\,  \ref{SampleSelection} we detail the data used for this study. In \S\,  \ref{MZgama} we introduce the M-Z relation for GAMA, and present the M-SFR and M-sSFR in \S\, \ref{MSFRgama}. In \S\, \ref{Environment} we discuss the environment influence on these relations and in \S\, \ref{Simulations} we apply a similar procedure to galaxies from a cosmological numerical simulation. Finally, in \S\,  \ref{Conclusion} we present a summary of our findings.

\section{Sample selection}\label{SampleSelection}

\begin{table*}
\centering
\caption{For every volume limited sample: redshift median and range,  magnitude limit, number of galaxies and mass median and range (95\% highest density interval)}
\begin{tabular}{lccrrrc}
\hline
\hline
Volume  & Redshift range & Redshift median & Magnitude limit (M$_r$)& \# of Galaxies &  Median [$\mathrm{log (M_{\odot})}$] & Mass HDI95\% [$\mathrm{log (M_{\odot})}$]\\ \hline
V1 & 0.04 - 0.13 & 0.0854 & -19.1 & 352 & 9.71 & 9.07 - 10.57\\
V2 & 0.13 - 0.225 & 0.1836 & -20.4 & 299 & 10.06 & 9.59 - 10.76\\
V3 & 0.225 - 0.36 & 0.2925 & -21.6 & 105 & 10.64 & 10.17 - 10.11\\ \hline
\end{tabular}
\label{Vsamples_tab}
\end{table*}

GAMA is a spectroscopic survey using data taken with the 3.9m Anglo-Australian Telescope (AAT) using the 2dF fibre feed and AAOmega multi-object spectrograph \citep{Sharp06}, the spectra were taken with  2 arcsec diameter fibres, a spectral coverage from 3700 to 8900 {\AA}, and spectral resolution of 3.2 {\AA}. For further details see  \citet{Baldry10, Driver11, Hopkins13, Liske15}.

GAMA has surveyed a total of $\sim$ 286 deg$^2$ split into five independent regions; three equatorial (called G09, G12, and G15), and two southern (G02, G23) fields. GAMA-I refers to a subset of data from the equatorial regions, and GAMA-II to the full five regions, see \citet{Liske15} for further details. In this paper, we are using GAMA-II data for the three equatorial regions.  For the equatorial regions G09, G12 and G15, spectra and redshifts are available for a high  redshift completeness of 98.48\%
of the galaxies within $r<19.8$  \citep[][]{Liske15}.

Galaxies in groups and clusters were selected from the GAMA Galaxy Group Catalog (G$^3$C) described in \cite{Robotham11}. The G$^3$C catalog uses a Friends-of-Friends (FoF) algorithm, which has been extensively tested on semi-analytic derived mock catalogues \citep[see also][]{Merson13}, and has been designed to be extremely robust to the effects of outliers and linking errors. In our analysis, a lower bound of groups with at least six galaxies has been set (this allows us to calculate the projected galaxy number density $\Sigma_5$ to study the local environment).

The spectroscopic data used in this paper are taken from the SpecLineSFRv05 GAMA catalogue, which includes the equivalent widths and Gaussian line flux measurements for the most prominent emission lines in GAMA-II spectra. For further details see \citet{Gordon17}. Emission line measurements are in agreement with an earlier emission line catalogue from GAMA-I \citet{Hopkins13} using the Gas AND Absorption Line Fitting algorithm \citep[GANDALF, ][]{Sarzi06}.

All emission line fluxes were extinction corrected using the Balmer decrement, assuming Case B recombination  \citep[][]{Osterbrock89}, {H$\alpha$}/{H$\beta$} $= 2.86$, and the extinction law of \citet{Cardelli89}. When {H$\alpha$}/{H$\beta$} $<2.86$ no correction is applied.



\begin{figure}
\centering
\includegraphics[width=\columnwidth]{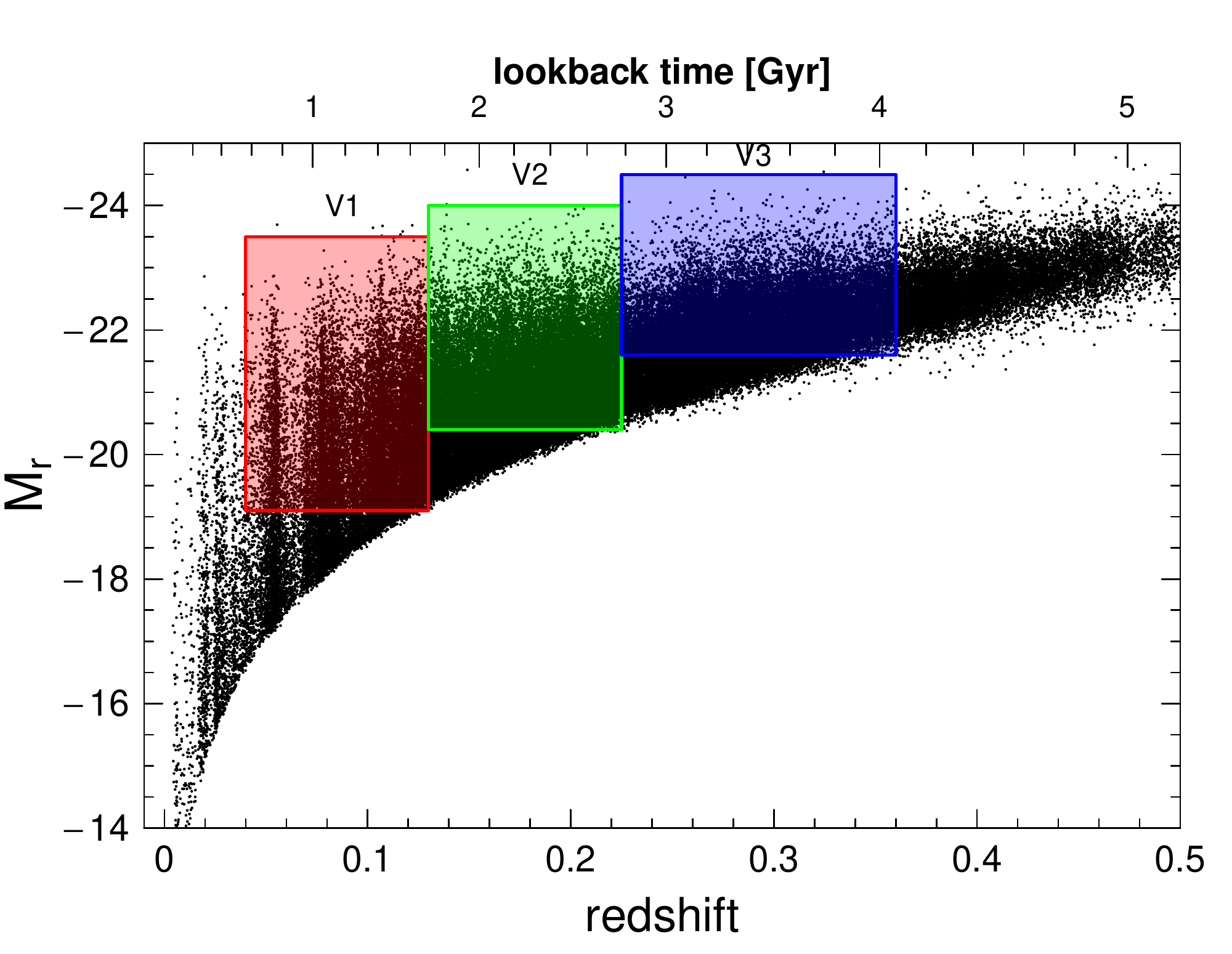}
\caption{Selected volume-limited samples for the GAMA survey. Values for the redshift and magnitude limits are shown in Table \ref{Vsamples_tab} }
\label{Vsamples}
\end{figure}

\begin{figure}
\includegraphics[width=\columnwidth]{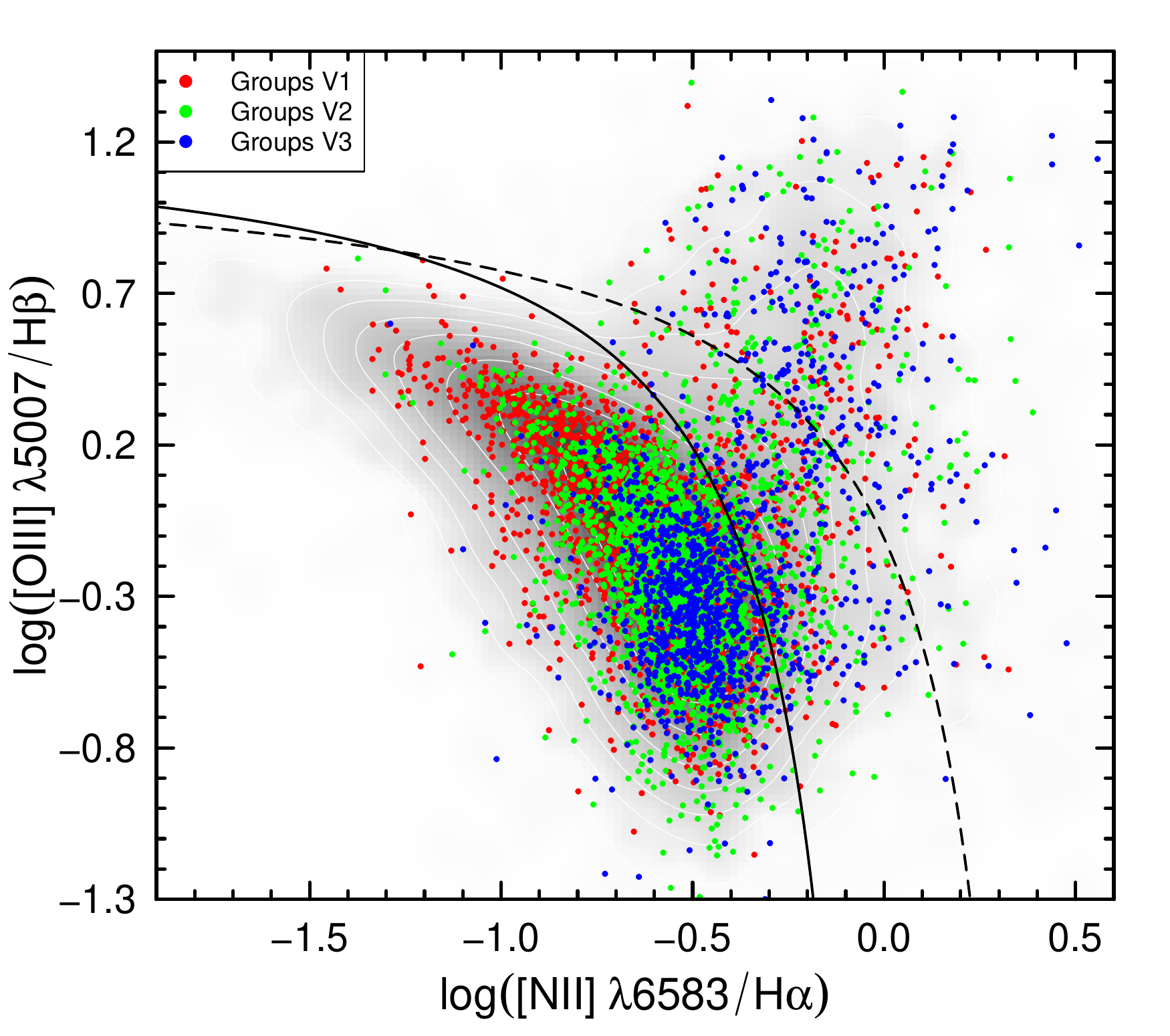}
\caption{BPT Diagram for the GAMA spectroscopic sample. The solid line shows the \citet{Kauffmann03} empirical division between SF and Composite galaxies, and the dashed line represents the \citet{Kewley01} starburst limit. Red, green and blue dots represent group galaxies in the volume-limited samples V1, V2 and V3, respectively. The gray background shows the full GAMA sample.}
\label{BPTDiagram}
\end{figure}

To avoid biases in volume and evolutionary effects due to redshift, we constructed 3 volume-limited samples (V1 to V3), by setting limits in M$_r$ and redshift as indicated in  Table \ref{Vsamples_tab} and Fig. \ref{Vsamples}.
This will not alter the multiplicity (total number of galaxies per group from G$^3$C) of a given galaxy, but can imply that some of the galaxies are excluded from further statistical analysis. For reliable metallicity and SFR estimates, we selected galaxies with a signal-to-noise ratio (SNR) of 3 in  {H$\alpha$}, {H$\beta$}, [\ion{O}{III}]$\lambda$5007 and  [\ion{N}{II}]$\lambda$6583. We selected star-forming galaxies using the BPT diagram \citep[][Fig. \ref{BPTDiagram}]{Baldwin81}, \mbox{and} the criteria of \citet{Kauffmann03}. Galaxies classified as Composite and AGNs were selected following the criteria of \cite{Kewley01}.  The total number of SF, Composite and AGN galaxies in each volume limited sample is listed in Table \ref{VolumeTable}. There are a total of 756 group galaxies classified as SF in the final sample.

\begin{figure}
\centering
\includegraphics[width=\columnwidth]{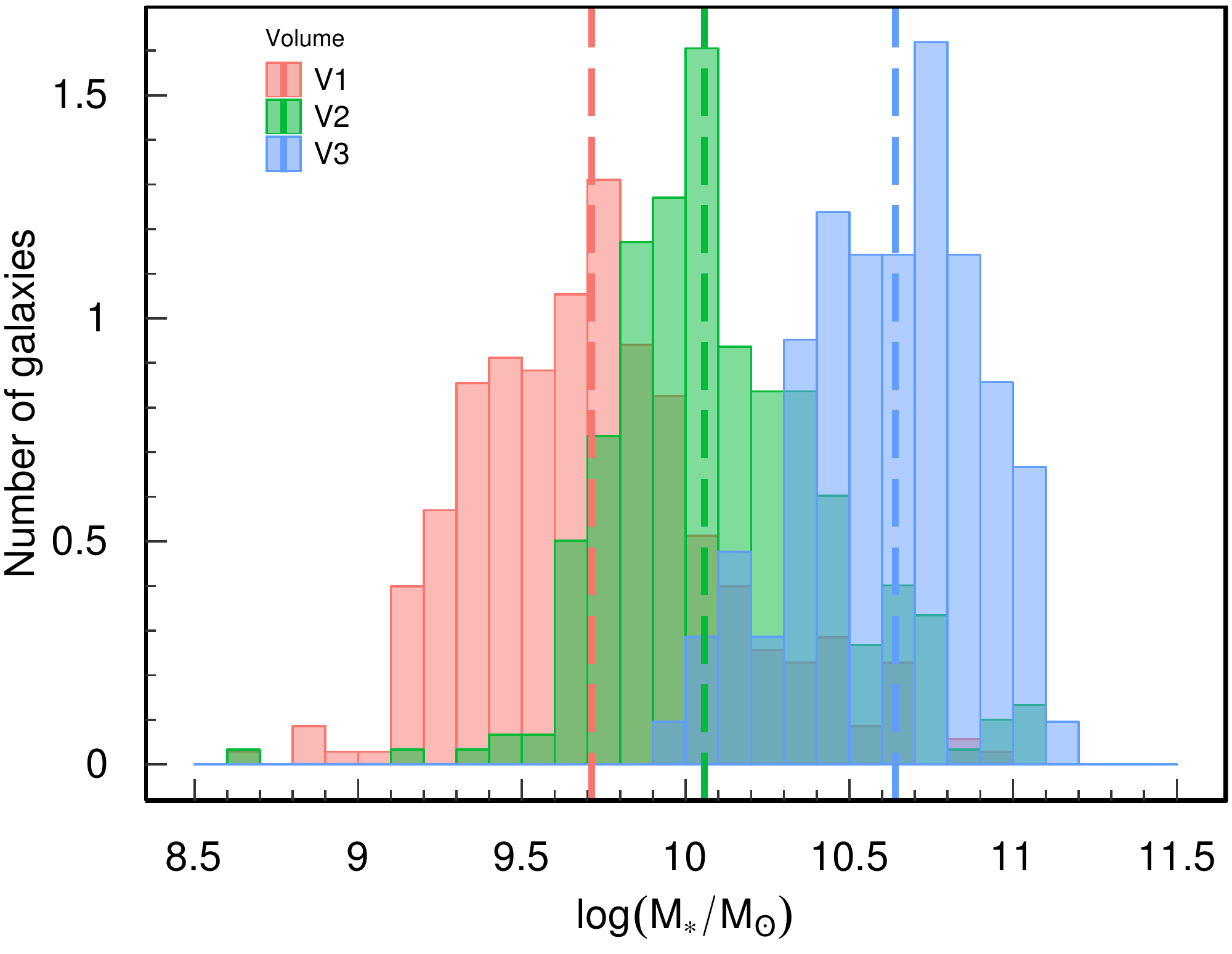}
\caption{Mass histograms for all three volume limited samples. Same color coding as in Fig. \ref{BPTDiagram}. Dashed lines represent the median of each volume sample.}
\label{histMstar}
\end{figure}

Stellar masses were estimated  by  \citet{Taylor11} (StellarMassesv19 GAMA catalog),  who estimate the stellar mass-to-light ratio ($M_{\ast}$/L) from optical photometry  using  stellar population synthesis based on the \citet{BC03} models. Figure \ref{histMstar} shows the mass distribution for each volume sample.

Metallicities were estimated only for SF galaxies using the extinction corrected fluxes, and the empirical calibration  provided by  \citet{Pettini04}  between the oxygen abundance and the O3N2 index:

\begin{equation}\label{eqn:O3N2}
\rm{O3N2} \equiv log \left(\frac{ \left[  OIII \right] \lambda5007\,  / H\beta}{ \left[ NII \right] \lambda6583 \, / H\alpha}\right).
\end{equation}

Finally, metallicities were recalibrated to the Bayesian system of \citet{Tremonti2004} using the calibration of \citet{Lara13}.

Star formation rates (SFR) were estimated following \citet{Gunawardhana11}, using the  equivalent width (EW) of H$\alpha$ to estimate the luminosity, correcting for aperture effects, obscuration, and stellar absorption.

\begin{table}
\centering
\caption{Total number of star-forming, AGN and Composite galaxies}
\begin{tabular}{lrrrr}
\hline
\hline
Volume & SFG & Composite & AGN & Total \\ \hline
V1 & 2298 & 215 & 100 & 2613 \\
V2 & 2123 & 298 & 118 & 2539 \\
V3 & 933 & 222 & 159 & 1314 \\
\hline
Total  & 26174 & 2556 & 1343 & 30073 \\ \hline
\end{tabular}\label{VolumeTable}
\end{table}


To detect any possible change in the metallicity and SFR of group galaxies, a control sample of field galaxies was constructed for each volume-limited sample with the same redshift, $r$-band magnitude, and stellar mass ranges,
following a similar approach to \citet{Kewley06} and \citet{Ellison09}.

\begin{figure}
\centering
\includegraphics[width=\columnwidth]{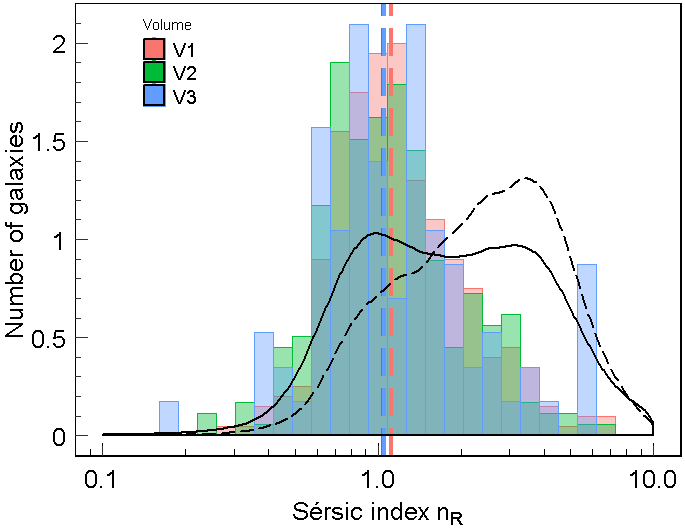}
\caption{{S\'ersic index histogram for the R-band. The volume-limited samples V1, V2 and V3 for star-forming galaxies are shown in red, green and blue, respectively. The solid contour shows the distribution for the whole GAMA sample, while the dashed contour for all galaxies in groups.}}
\label{histSersic}
\end{figure}

Since we aim to analyse the gas metallicity of group galaxies, our sample is composed of SF galaxies with \mbox{emission} lines, biasing our sample to late-type galaxies as seen in the S\'ersic index distribution of Fig. \ref{histSersic}. The same figure shows the distribution for the whole GAMA sample, and for all group galaxies. As expected, group galaxies in general show a higher proportion of early-type morphologies, consistent with  \citet{postman84, dressler97,  postman05, bamford07, calvi12}, among others.  

\begin{figure}
\centering
\includegraphics[width=\columnwidth]{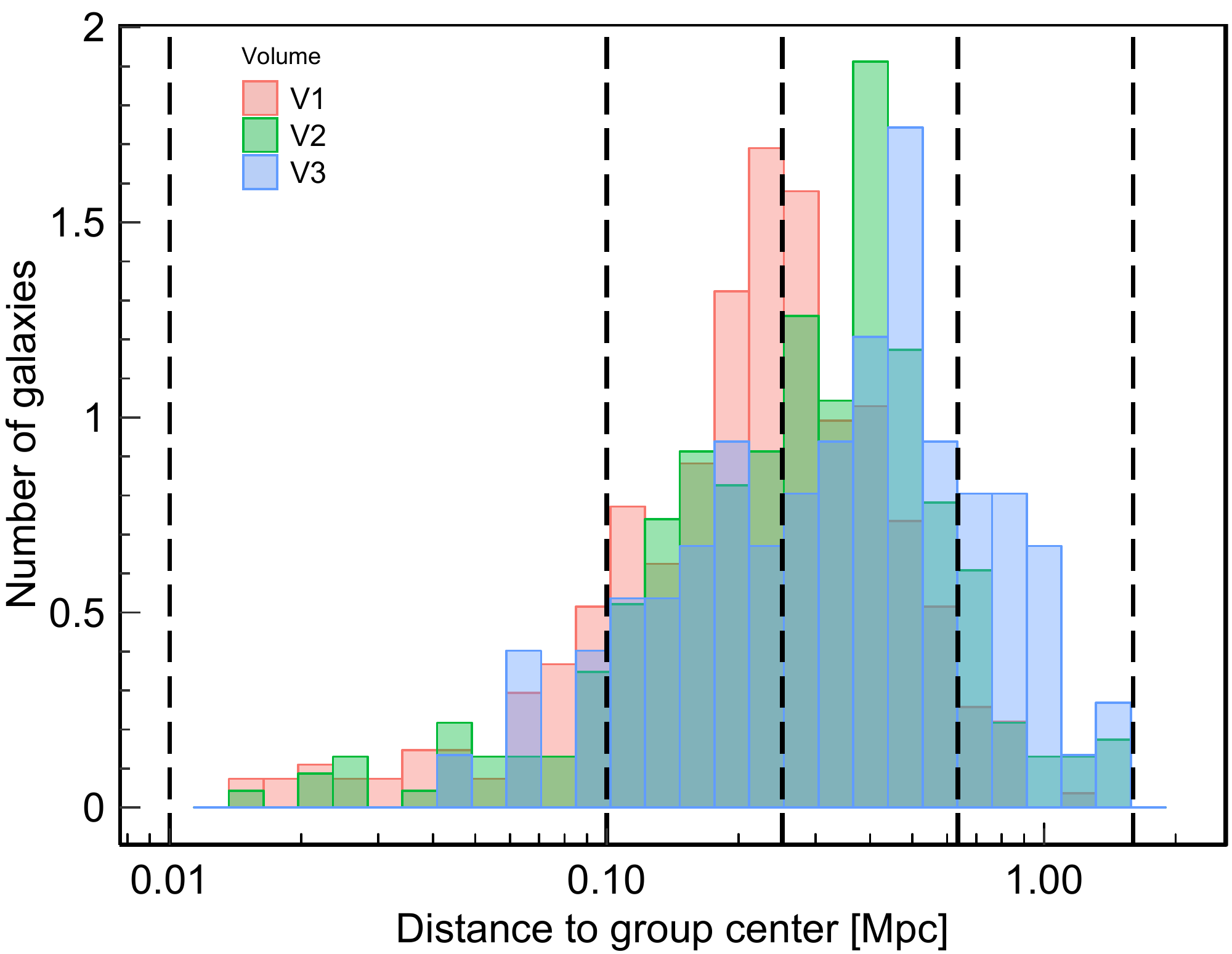}
\caption{Distances to group center. Dashed lines represent the limits of the four ranges used in section \ref{Environment}.}
\label{histDistIterCen}
\end{figure}

\begin{figure}
\centering
\includegraphics[width=\columnwidth]{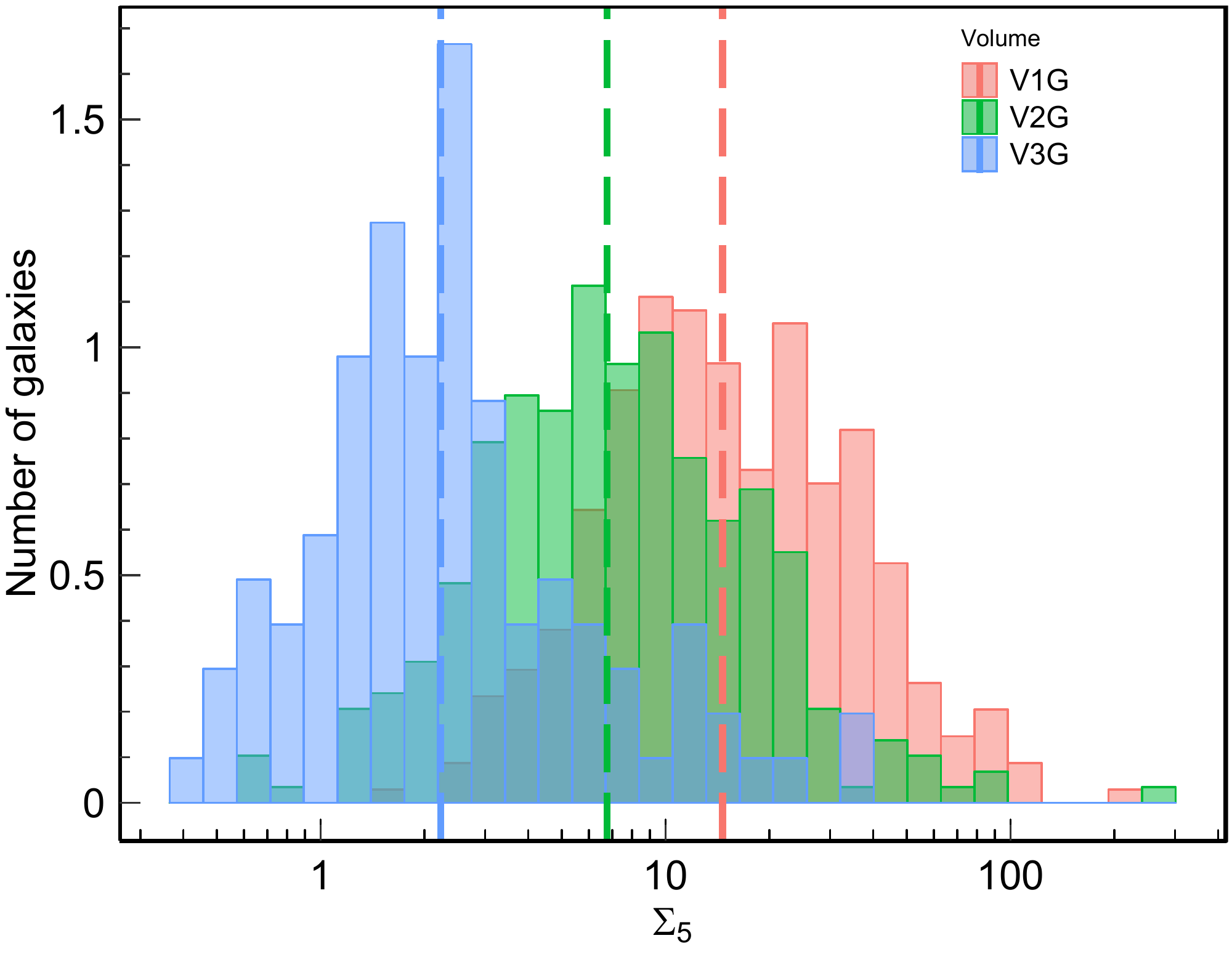}
\caption{Distribution of $\Sigma_5$  for all three volume limited samples. Same color coding as in Fig. \ref{BPTDiagram}. Dashed lines represent the medians of each volume limited sample.}
\label{histS5}
\end{figure}

{Finally, to further characterise our sample, the histogram in Fig. \ref{histDistIterCen}  shows the distribution of galaxy distances to the group center for our three volume-limited samples. To quantify the influence of local environment (see section \ref{Environment}), we estimated the surface number density of \mbox{galaxies} $\Sigma_5$, as defined by \citet{Muldrew2011}. The area used to calculate $\Sigma_5$ was that of the circle with radius equal to the projected distance to the fifth nearest neighbor galaxy. The final distribution of $\Sigma_5$ is shown in Fig. \ref{histS5}.}

{It is worth noting that our final sample is formed only by SF galaxies, with at least four emission lines (\Ha, \Hb, [\ion{N}{ii}]$\lambda$6583 and [\ion{O}{iii}]$\lambda$5007) to estimate gas metallicities. This biases our sample to  late-type morphologies as indicated by their S\'ersic index. As a consequence, passive galaxies are not included in our sample. Moreover, as seen in Fig. \ref{histDistIterCen} our final sample has very few galaxies close to the group center. This is a result of the absence of passive galaxies, given the spatial distribution expected from morphology-density relation \citep[e.g.,][]{Dressler1980,Goto03}.}

\section{Scaling relationships for galaxy groups}\label{ScalingRelationsgama}

\subsection{Methodology}\label{Methodology}

To detect a reliable measurement of the enhancement or suppression of SFR and metallicity in group galaxies, we follow a similar methodology to \citet{Ellison08} and \citet{Garduno20} and generate control samples to be taken as a reference for each sub-sample of galaxy groups. To obtain reliable control samples to quantify any effect due to metallicity evolution \citep{Lara09a, Lara09b, Pilyugin11, Pilyugin13} and the intrinsic shape of the each scaling relation, it is important to compare each relation of group galaxies with its respective counterpart of control galaxies in the same redshift and stellar masses ranges.

To generate control samples, first we create a field galaxy catalogue by removing galaxies in pairs and groups from our main GAMA spectroscopic catalogue. This  results in a sample of $\sim$16,457 field galaxies. Second, for each galaxy in our group sample, we follow an iterative process that finds matches in redshift and stellar mass from the field galaxy sample. The iteration process finishes when the redshift and stellar mass distribution of the control sample matches the group sample (see inset histograms in Figs. \ref{MZ_V123} and \ref{MSFRSSFR_V123}). 

We create a control sample for each of the three volume-limited datasets 
described above (see also Fig. \ref{Vsamples_tab}). Since every volume limited sample spans different stellar mass ranges, in order to properly fit the shape of the scaling relationships, we created a fiducial fit to the 3 volume-limited samples together (V1+V2+V3).

\begin{figure*}
\centering
\includegraphics[scale=0.32]{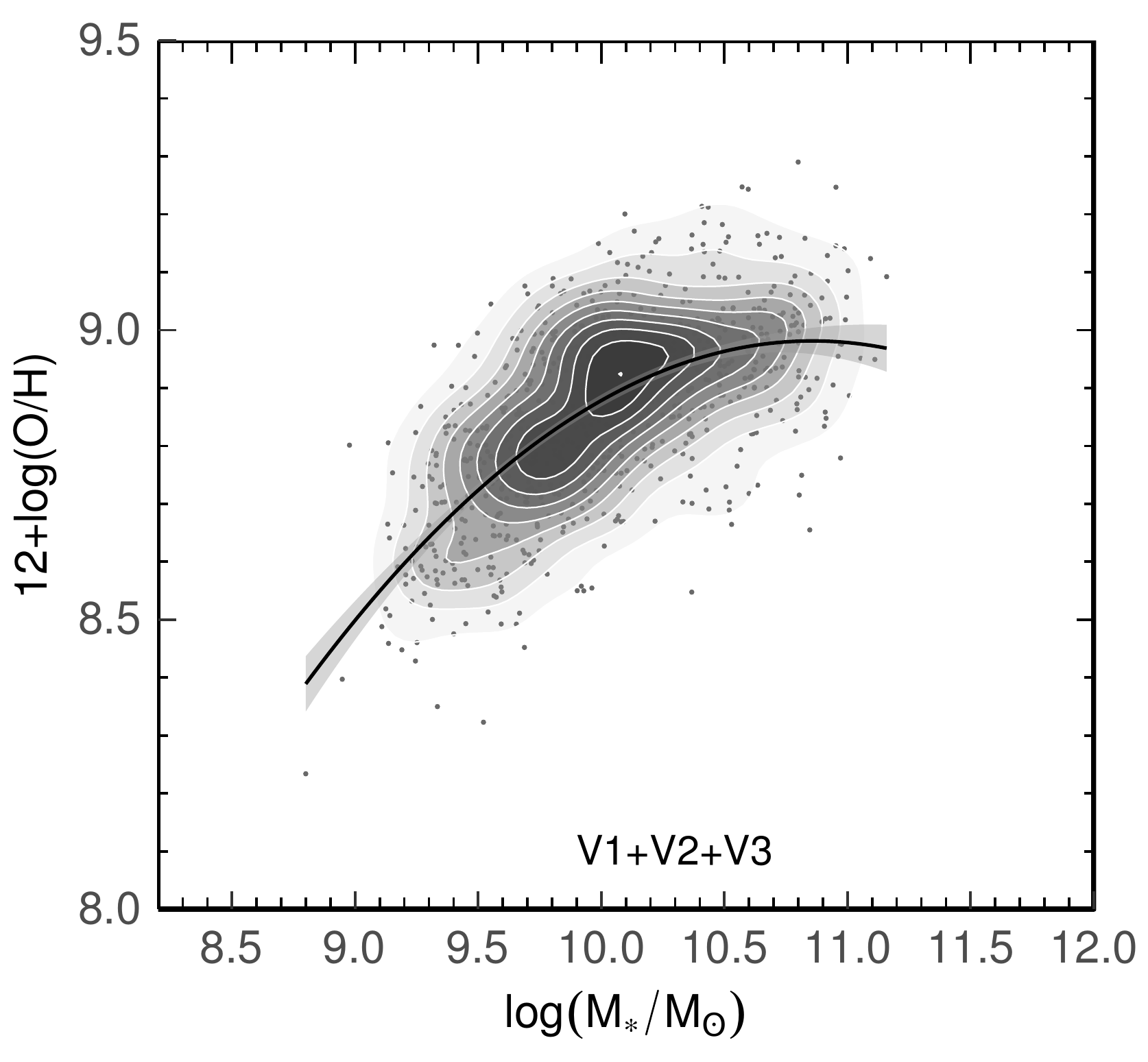}
\includegraphics[scale=0.32]{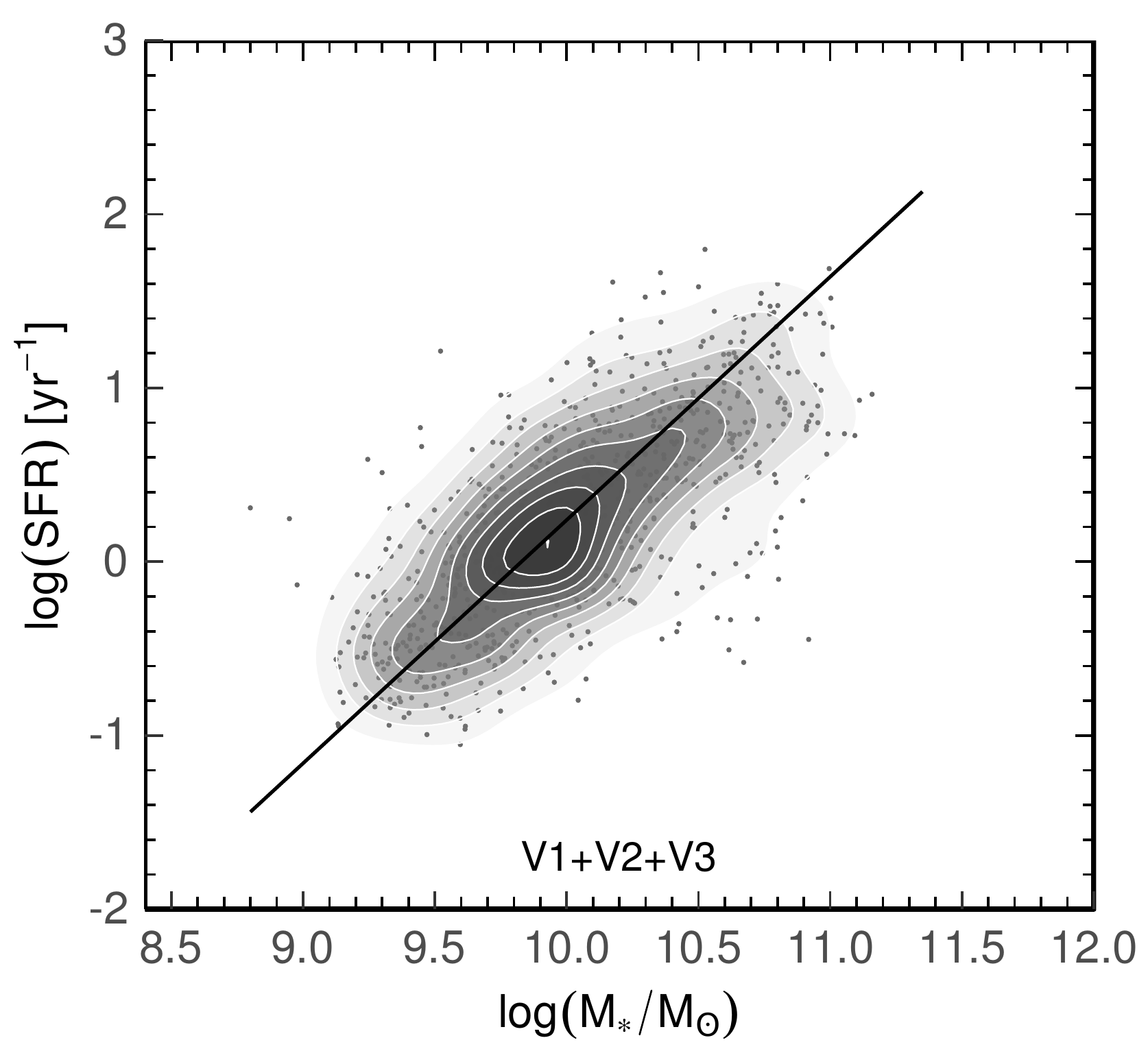}
\includegraphics[scale=0.32]{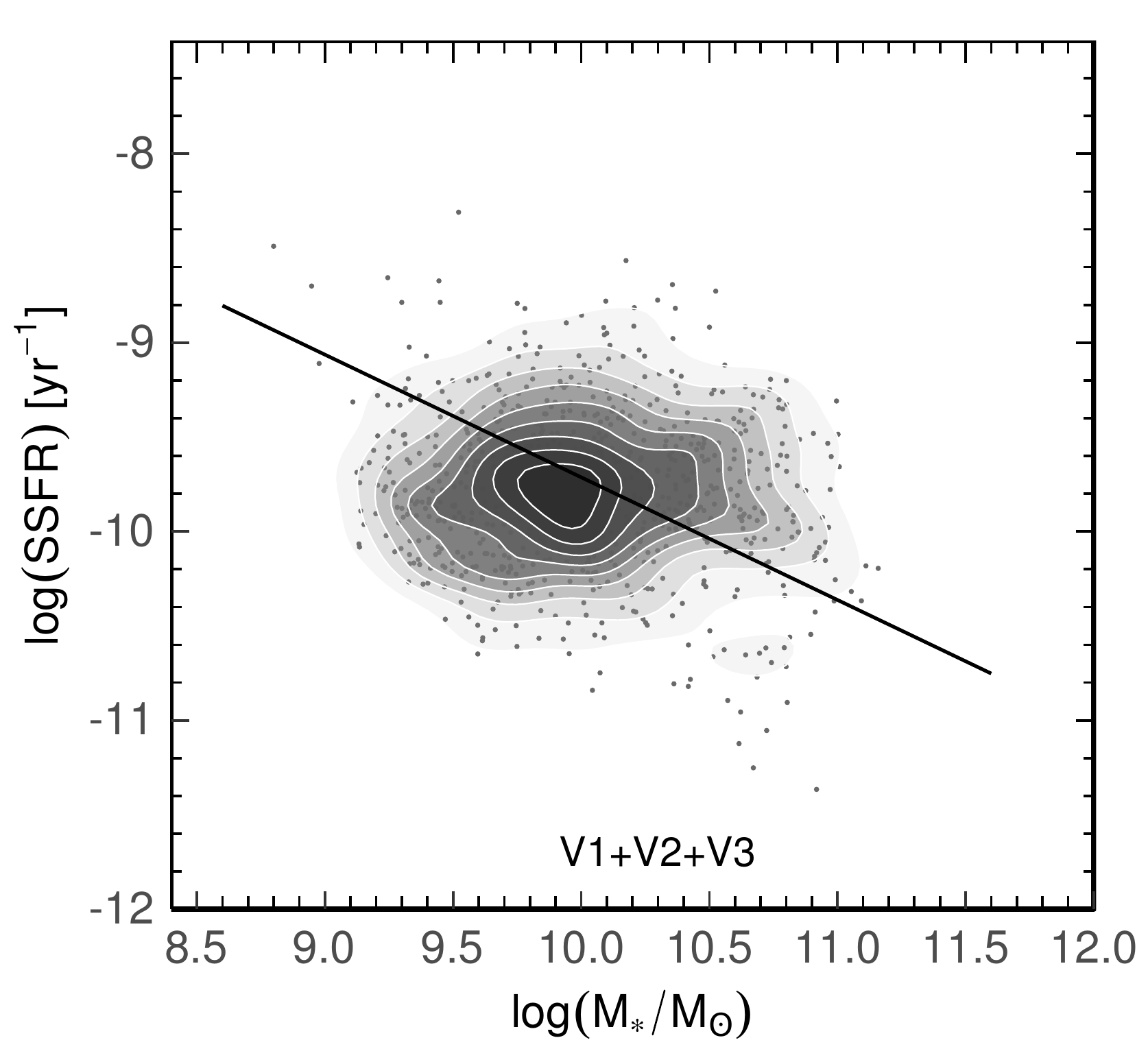}
\caption{M-Z (left), M-SFR (centre), and M-sSFR (right) relations for the joint control sample. Dots represent control galaxies for the V1, V2 and V3 volumes together, and the data density is represented with the shaded areas. The solid black lines are the best-fitting relations, or the fiducial fit, also shown in gray dashed lines in Fig. \ref{MZ_V123}.}
\label{MZSFRSSFR}
\end{figure*}

The M-Z, M-SFR and M-sSFR relations for the 3 joint control samples are shown in   Fig. \ref{MZSFRSSFR}. 
With the above argument in mind and following a similar approach to \cite{Lara13}, we fit the M-Z relation of the control sample for all galaxies in all three volume-limited samples using a second-order polynomial. The fitting is done by iteratively re-weighted least squares (IRLS).

We repeat the same proceedure for the M-SFR and M-sSFR using a one-order polynomial. The coefficients of the resulting fits are given in Table \ref{Vsamples_equ}.  For the one-order polynomials, the fitting is performed with the  hyper-fit routine in R \citep{Robotham2015}. Hyper-fit converts D-dimensional data with Gaussian uncertainties to a (D-1)-dimensional hyperplane with intrinsic scatter, using a maximum likelihood approach.


The fiducial fit to the joint set of the 3 volume-samples will function as a base fit, and will be used to measure offsets as indicated in the next section.


\begin{table}
\centering
\caption{From top to bottom, best-fit coefficients for the M-Z ($12+\log(O/H) = a + bx + cx^2$), M-SFR ($\log(sSFR) = a + bx$) and M-sSFR ($\log(sSFR) = a + bx$) relations. In all cases $x=\log(M_{\star}/\rm{M}_{\odot})$.}
\begin{tabular}{lrrr}
\hline
\hline
Relation & a & b & c \\ \hline
M-Z & -7.50 $\pm$ 1.76 & 3.04 $\pm$ 0.35 & -0.140$\pm$ 0.018 \\
M-SFR & -13.95 $\pm$ 0.47 & 1.415 $\pm$ 0.046 & - \\
M-sSFR & -4.22 $\pm$ 0.79 & -0.555$\pm$ 0.079 & -  \\
\hline
\end{tabular}
\label{Vsamples_equ}
\end{table}

\subsection{The M-Z relation for galaxy groups}\label{MZgama}

\begin{figure*}
\centering
\includegraphics[scale=0.34]{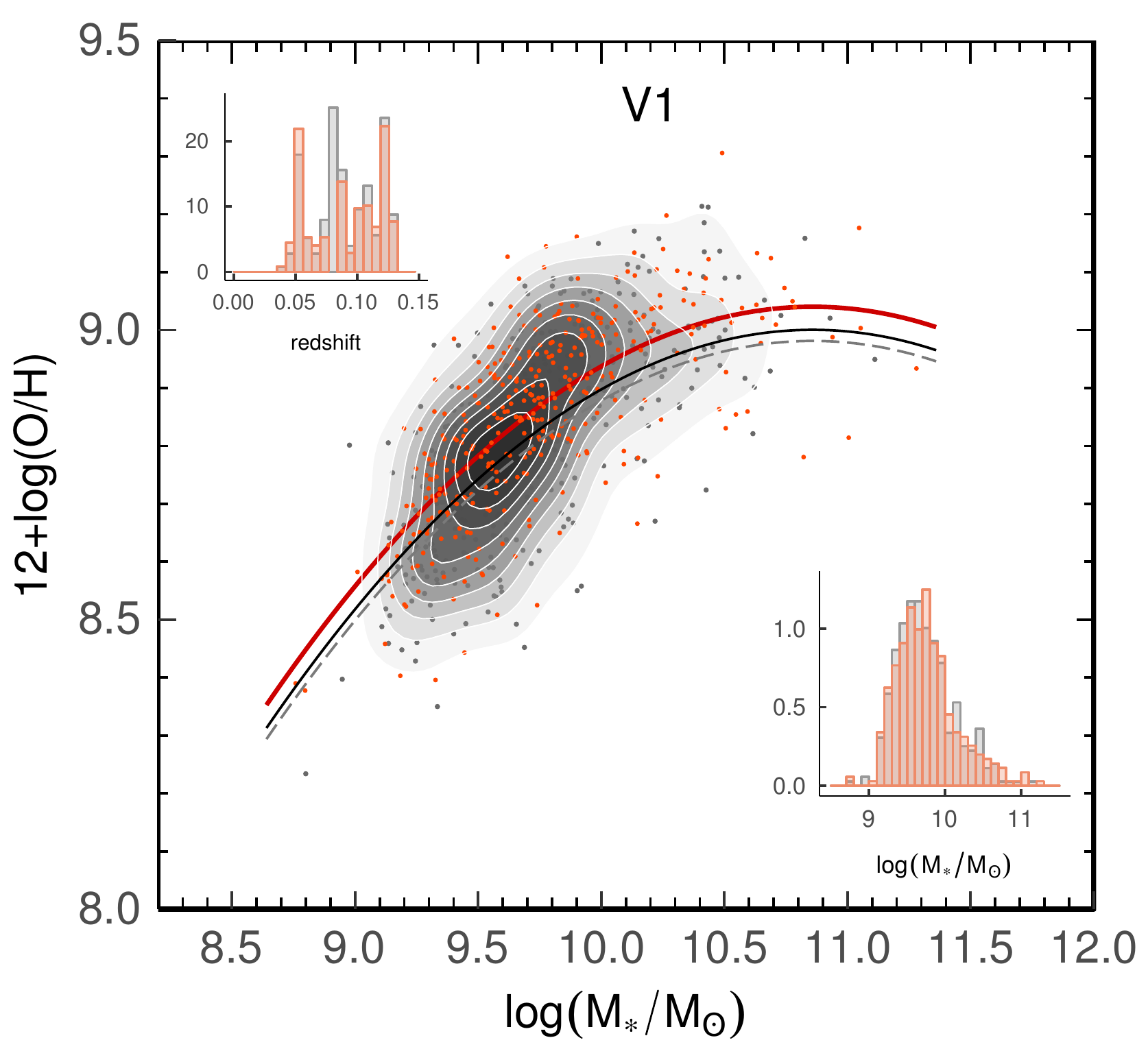}
\includegraphics[scale=0.34]{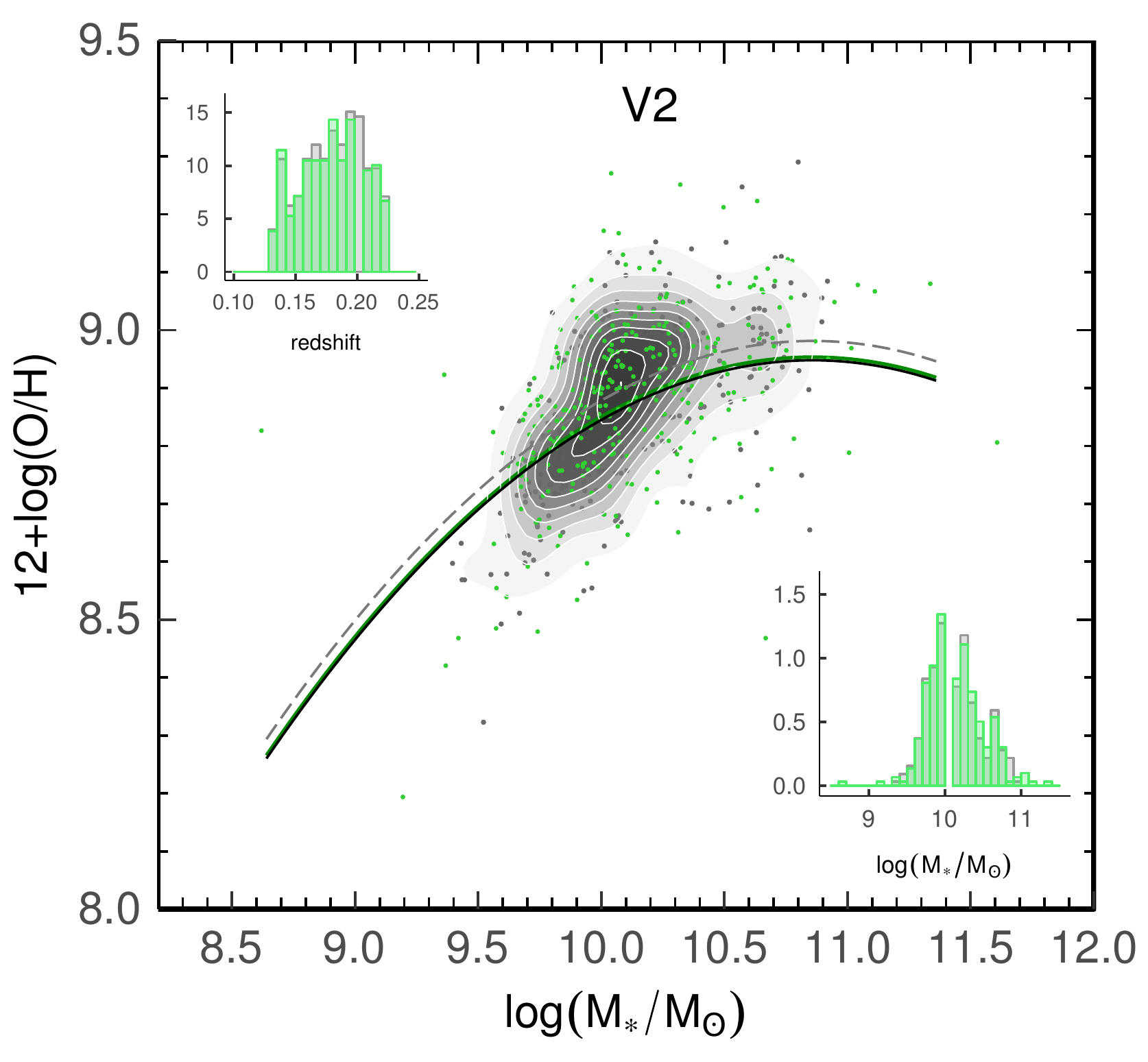}
\includegraphics[scale=0.34]{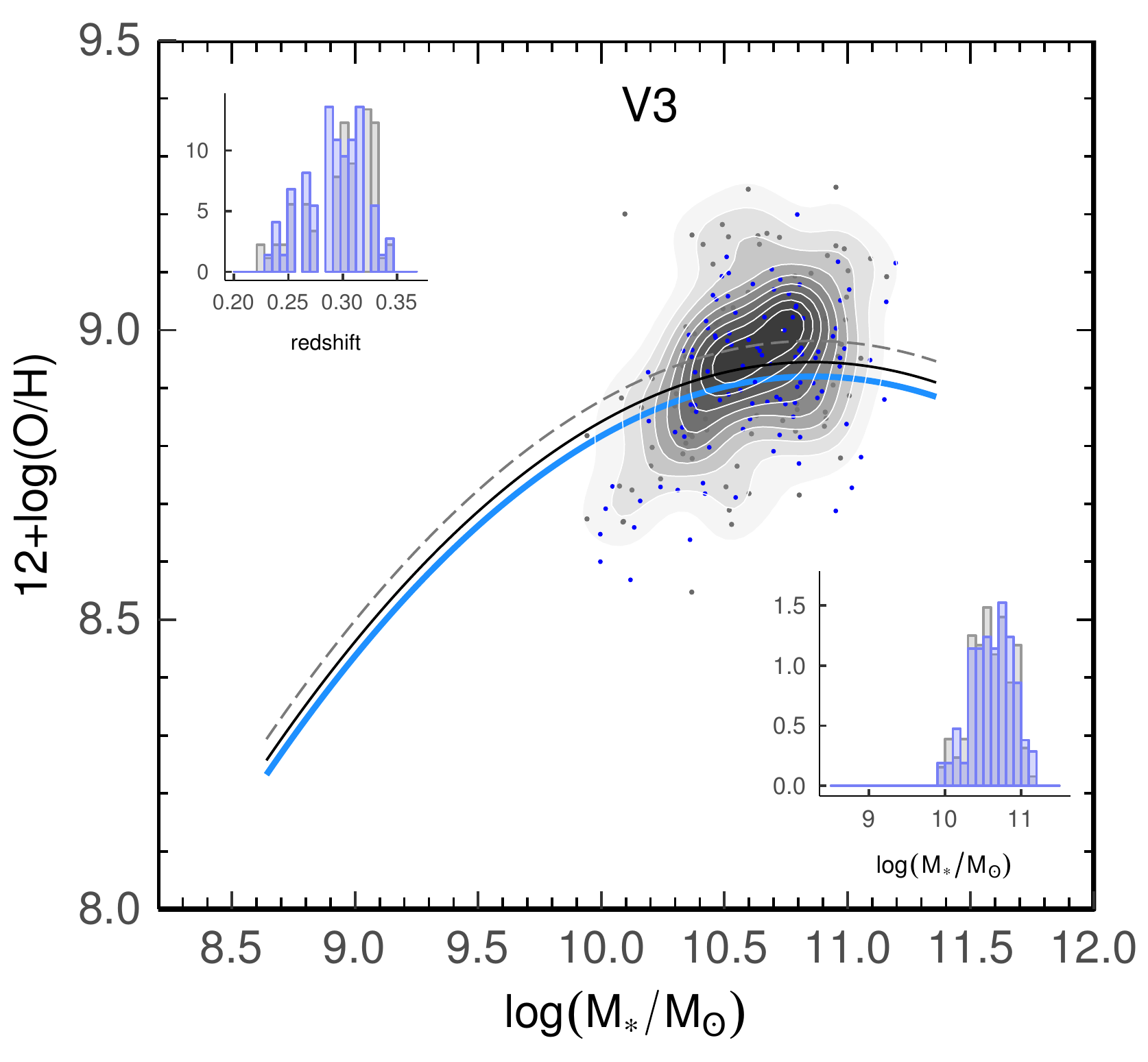}
\caption{M-Z relations for our three volume limited samples V1, V2, V3 (left, centre, right, respectively). Grey dots represent control galaxies (with the shaded areas as the data density), while coloured dots denote group galaxies. Red colours (dots, solid line fit, and histogram), correspond to galaxies in groups in the volume limited sample V1; similarly, green for V2, and blue for V3. The  black solid line represents the best fit for the control sub-sample, whereas the gray dashed line is the fiducial fit to all control samples. Inset histograms represent the redshift (upper) and mass (lower) distributions of the groups (coloured) and control (gray) samples.}
\label{MZ_V123}
\end{figure*}

In this section we use the M-Z relation for the three joint control samples (described above) as a baseline. Next, we proceed to fix the $b$ and $c$  coefficients of the fiducial fit for the M-Z relation, and then fit the zero point $a$ separately for the group and the control galaxies for each volume-limited sample, as shown in Fig. \ref{MZ_V123}. As indicated in the same figure, the coloured line corresponds to the groups, the black solid line to the control samples, and the dashed gray line is our fiducial fit.


From the fitting procedure described above, we define the difference $\Delta$Z = a$_{\rm group}$ - a$_{\rm control}$, as the difference in the fitted zero points of the M-Z relation. 


Confidence intervals for the differences of the zero point coefficients are calculated using bootstrapping. We create 1020 artificial sub-samples through random selection with replacement from the original samples, using the same fitting technique to estimate the coefficient $a$ as described above, with the range of offset found taken to be the uncertainty in the measurement.
The offsets in $\Delta$Z are shown in Figure \ref{diffV} and their values together with the uncertainties are given in Table  \ref{diffVsamples_tab}.




It is important to take into account that each of the samples V1 to V3 are sampling different ranges of stellar mass, since their member galaxies are selected in different magnitude limits (see Fig. \ref{histMstar} and Table \ref{Vsamples_tab}). For low redshift galaxies at low stellar masses, V1, our data suggests that group galaxies show a small increment of $\sim +0.04$ dex in metallicity with respect to the control sample (refer to table \ref{diffVsamples_tab} for errors). Our results for the local volume V1 are consistent with \citet{Ellison09}, who found $\sim +0.05$ dex higher gas metallicity for a sample of cluster galaxies in the SDSS survey.

On the other hand, for the intermediate volume V2, there is no  difference in gas metallicity between control and groups. More massive galaxies in groups at higher redshifts, V3, show a very small decrement of $\sim -0.024$ dex in their gas metallicity with respect to the control sample. These differences are shown as purple squares in \mbox{Fig. \ref{diffV}}. We highlight that since the total number of galaxies in the volume V3 are $\sim$100, the error bars for the offsets in this volume are larger, and they must be consider with caution.




\subsection{The M-SFR and M-sSFR relations for galaxy groups}\label{MSFRgama}

We now repeat the above analysis but for the relations between the stellar mass with SFR and sSFR for our samples.


Following the same procedure described above, we fix the $b$ coefficient of the M-SFR and M-sSFR relations, and fit the zero point $a$ separately for the group and the control galaxies of each volume-sample (see Fig. \ref{MSFRSSFR_V123}). As in the M-Z diagrams, 
the group and control population are shown for the three \mbox{volumes}, together with the best fit for M-SFR and M-sSFR in each. The fiducial fit of the joint set of the control sample and the distribution of mass and redshift for the different subsamples are represented as well. While the M-sSFR relation shows a large scatter, the derived fit is the most optimal.

\begin{figure}
\centering
\includegraphics[width=\columnwidth]{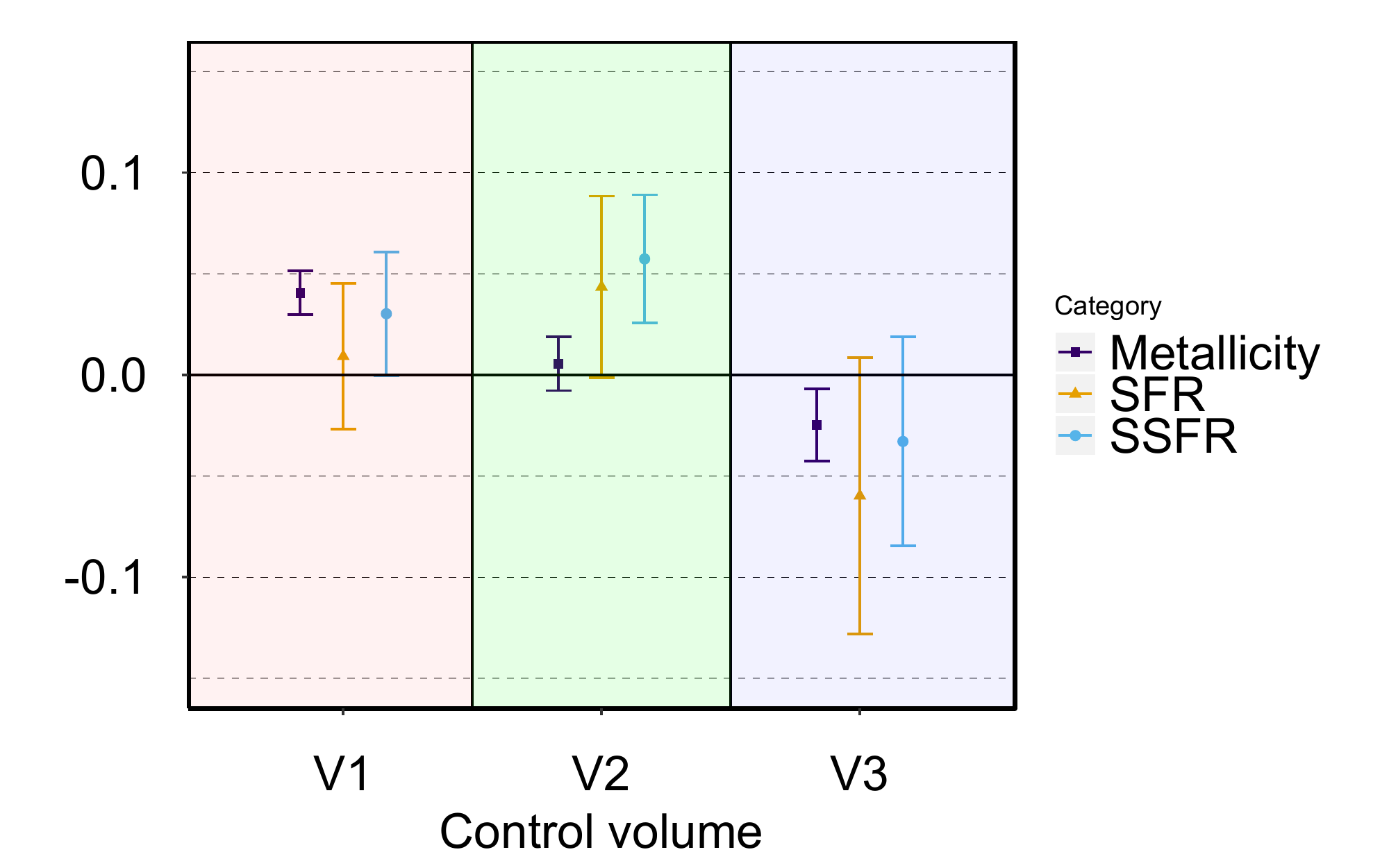}
\caption{Differences for the zero-point coeffient between groups and control galaxy samples, in metallicity, SFR and sSFR. Median and error bars of one standard deviation are represented}
\label{diffV}
\end{figure}

\begin{figure*}
\includegraphics[scale=0.34]{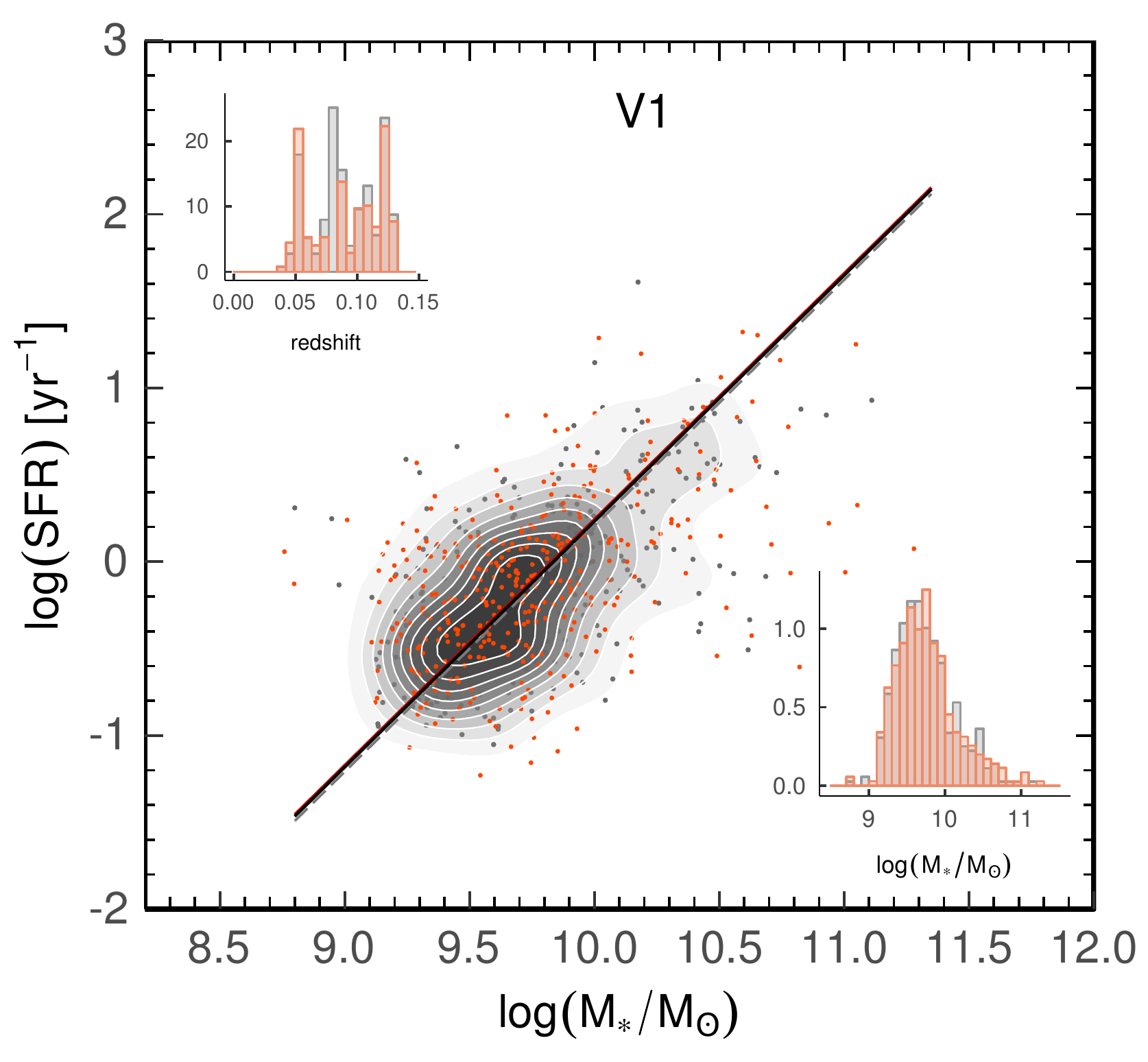}
\includegraphics[scale=0.34]{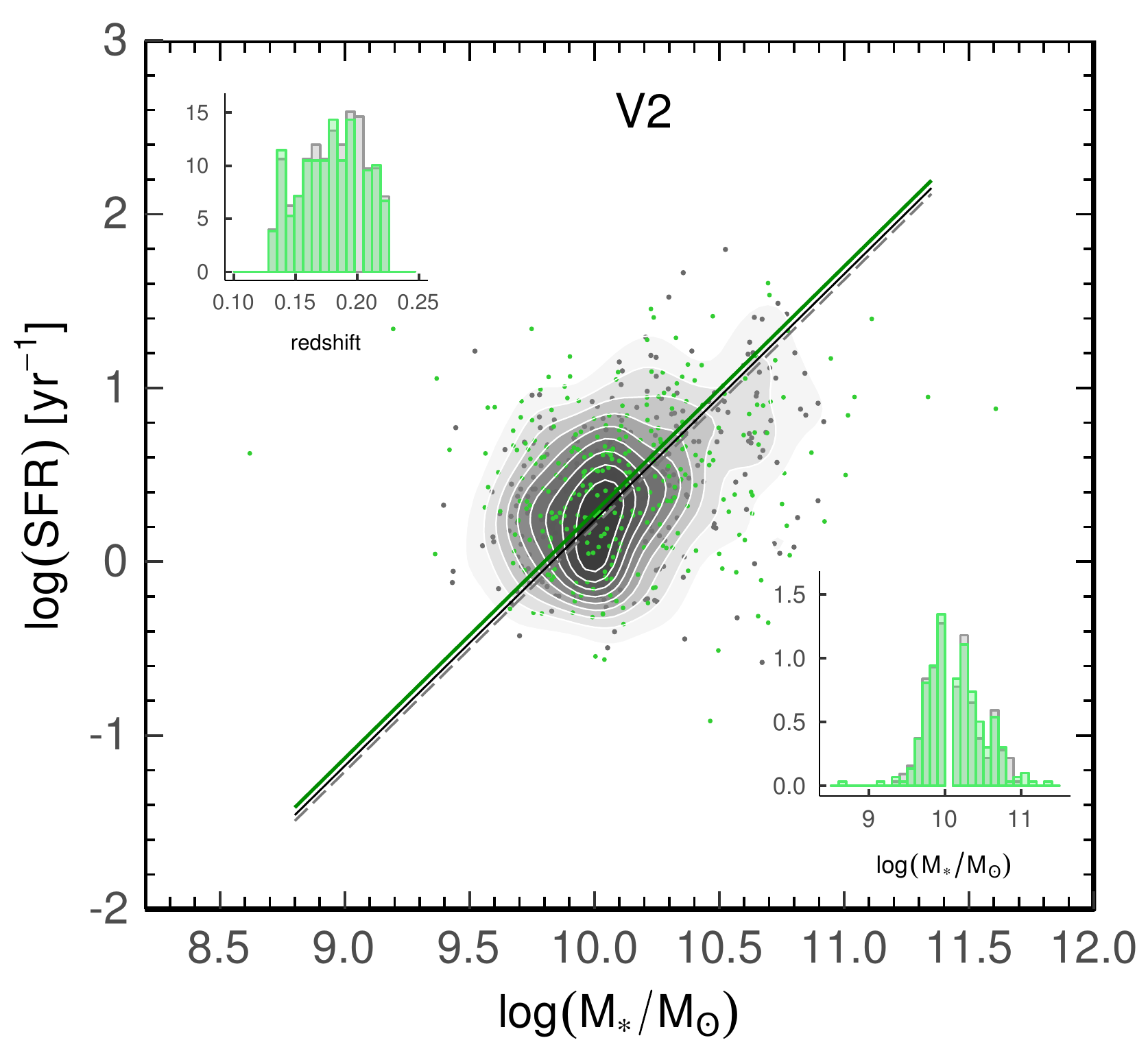}
\includegraphics[scale=0.34]{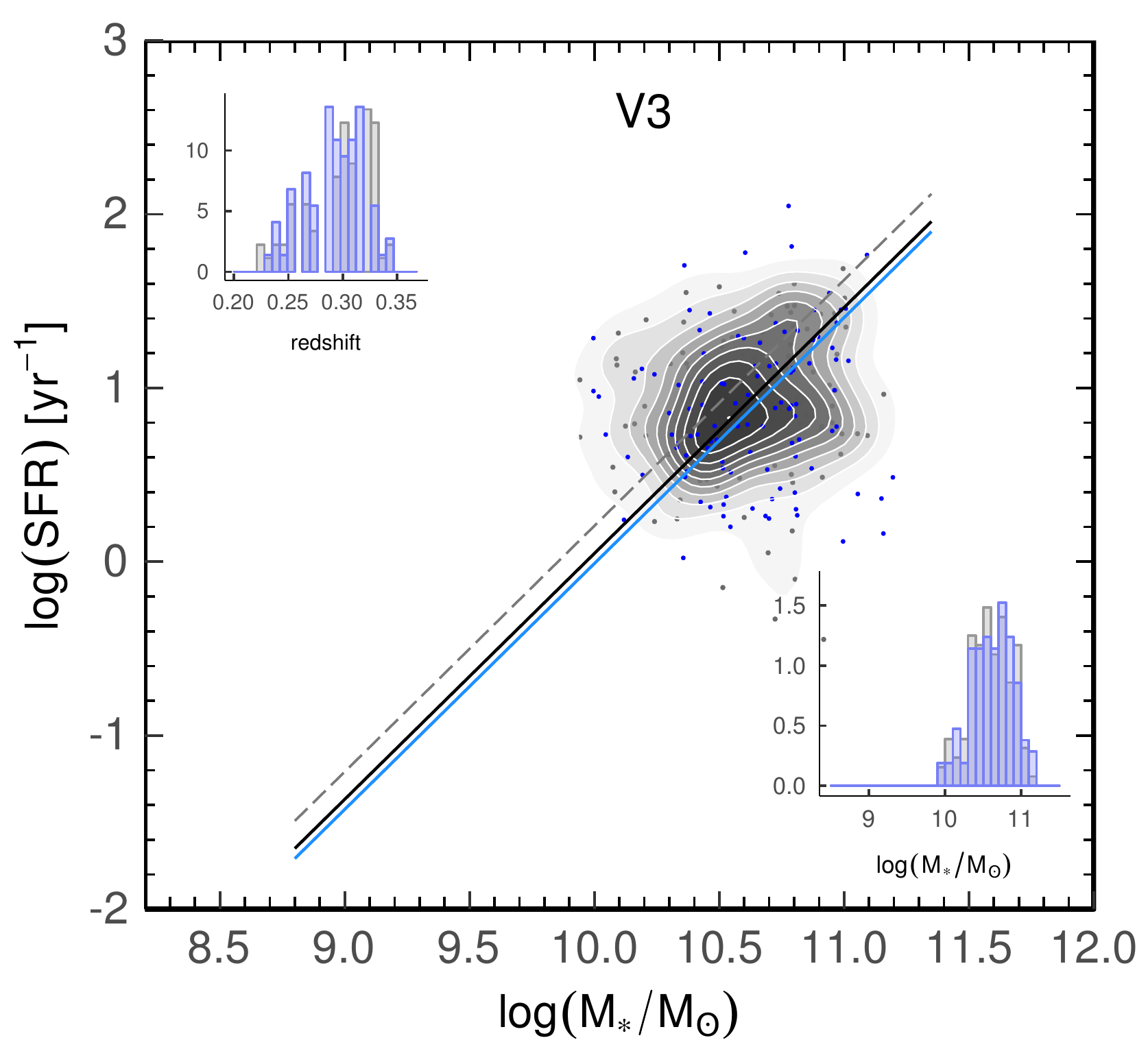}
\includegraphics[scale=0.34]{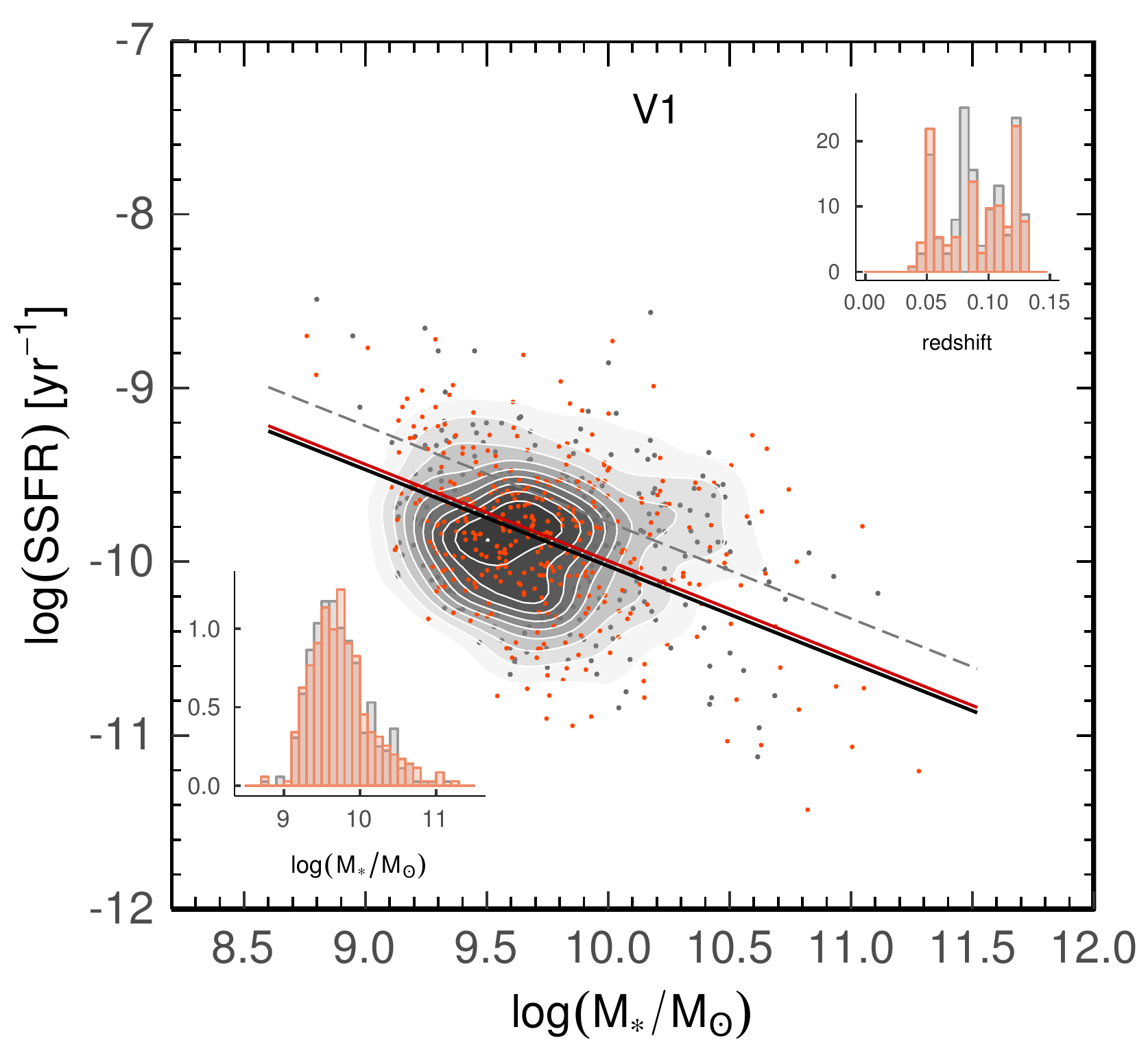}
\includegraphics[scale=0.34]{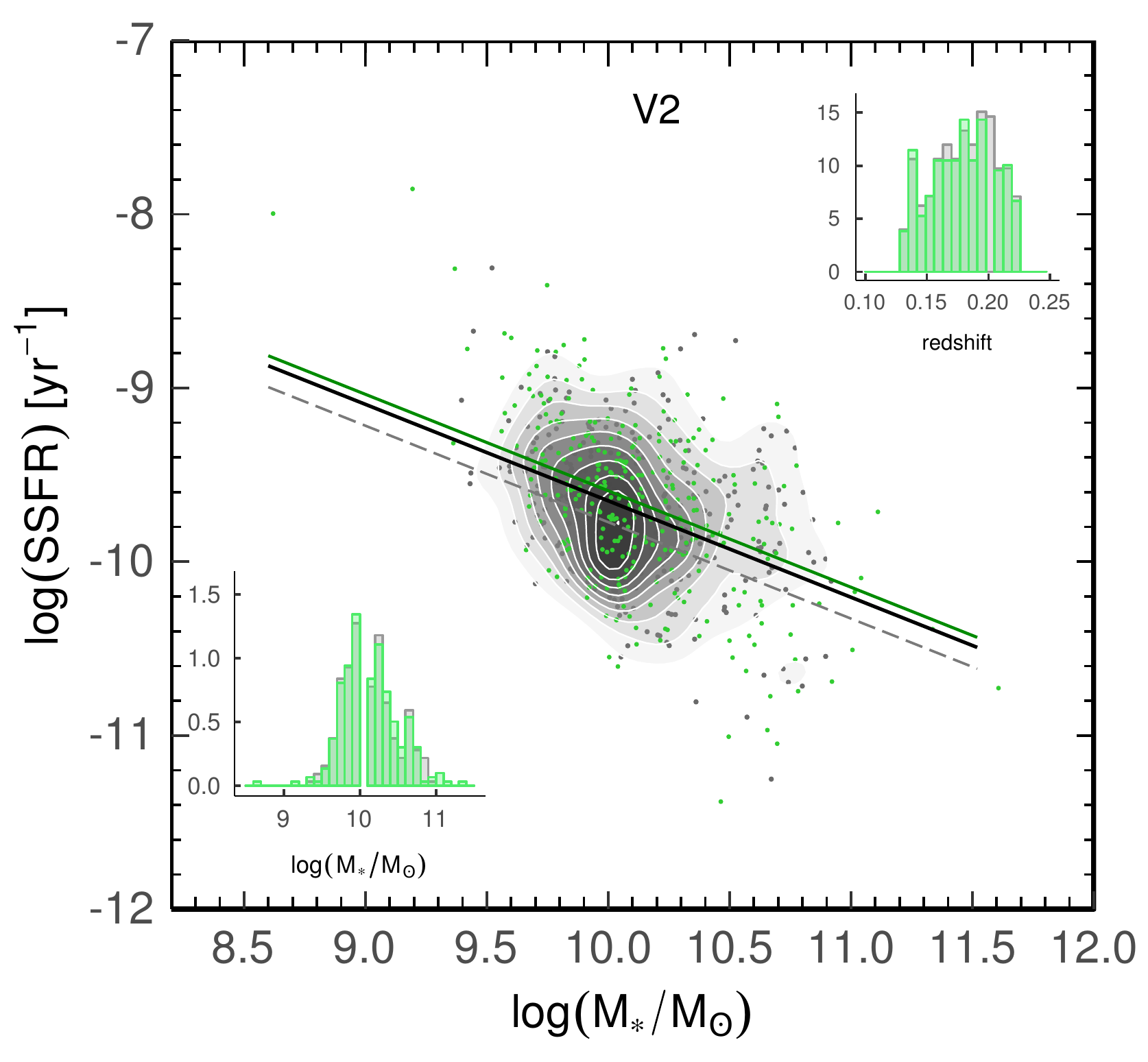}
\includegraphics[scale=0.34]{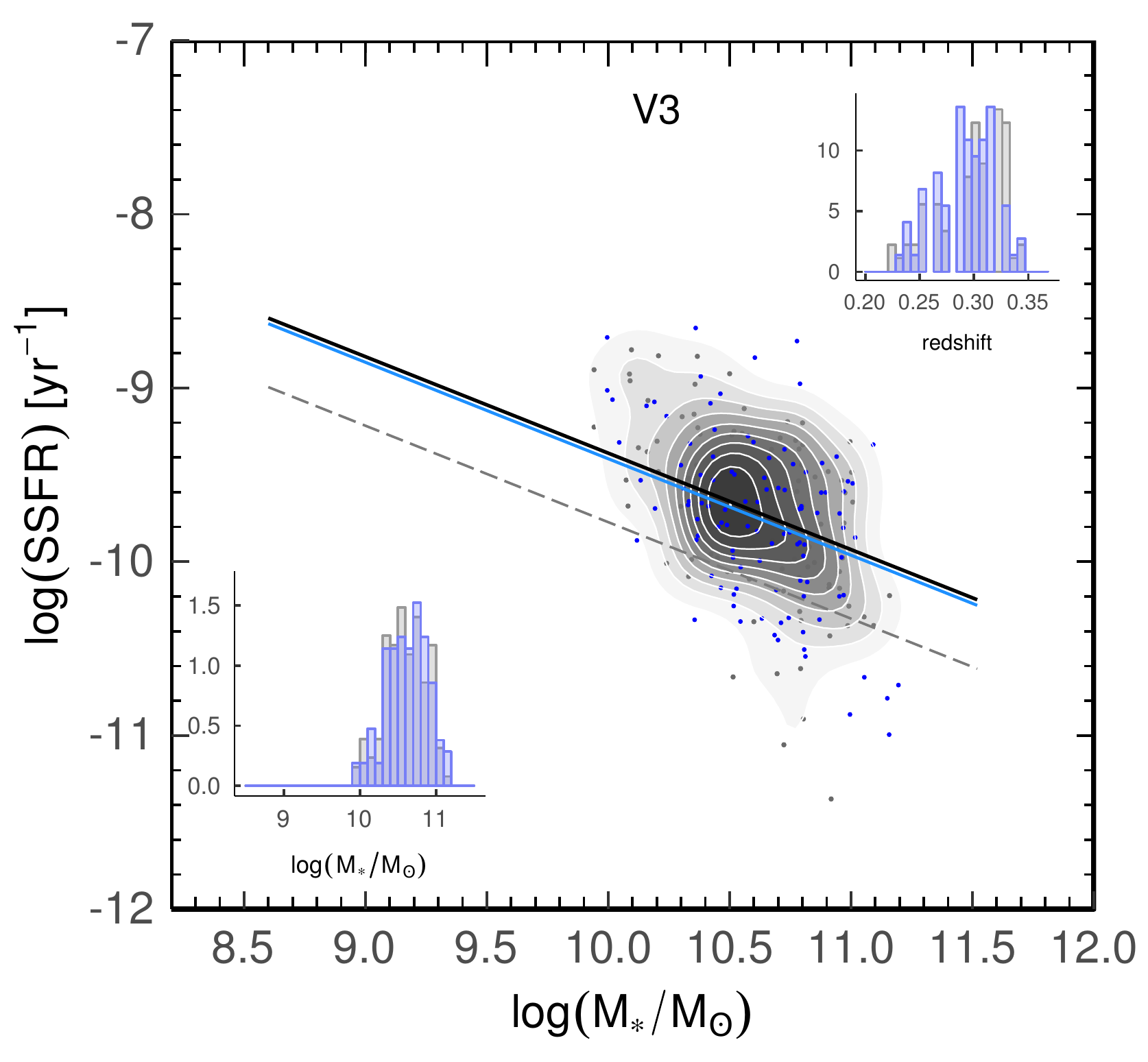}
\caption{M-SFR and M-sSFR relations for our three volume limited samples V1, V2, V3 (left, centre, right, respectively). Grey dots represent control group galaxies, coloured dots (red, green, blue for V1, V2, V3, respectively), group galaxies. The black line represents the best fit for the control subsample whereas the coloured line is the fit for the group galaxies subsample. The gray dashed line is the fiducial fit to the joint set of the control sample. Inset histograms represent the redshift (upper) and mass (lower) distributions of the groups (coloured) and control (grey) samples.}
\label{MSFRSSFR_V123}
\end{figure*}

Significant differences in the measured zero-point coefficient can be seen for SFR and sSFR in the different subsamples (yellow triangles and blue circles, respectively, in Fig. \ref{diffV}). For the V1 volume there is no \mbox{difference} for SFR, but sSFR is marginally higher in group galaxies with respect to field galaxies. In the intermediate range of redshifts of V2, there is a clear increment in both SFR and sSFR, whereas those (more massive) galaxies in groups at higher redshifts, V3, show a slight decrement in their SFR and sSFR with respect to field galaxies.

This is consistent with a scenario where the environmental mechanisms enhancing the star formation, mainly galaxy-galaxy interactions, are noticeable at the redshift range of V2. The processes quenching star formation (starvation, harassment), need longer times to produce observable effects and become dominant at the redshift of V1. 

Despite the larger uncertainties, the lower SFR and sSFR obtained for galaxies in the V3 volume (that are in general more massive as showed in Fig. \ref{histMstar}) agrees with previous work \citep{vonderLinden10,vulcani10,allen16}.
 
\begin{figure*}
\centering
\includegraphics[width=.66\columnwidth]{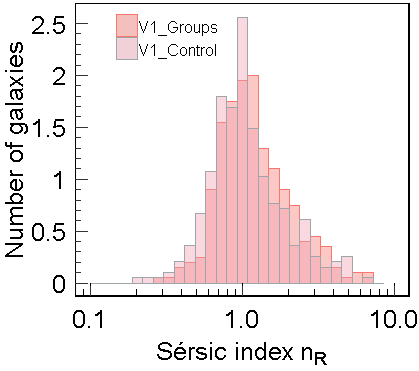}
\includegraphics[width=.66\columnwidth]{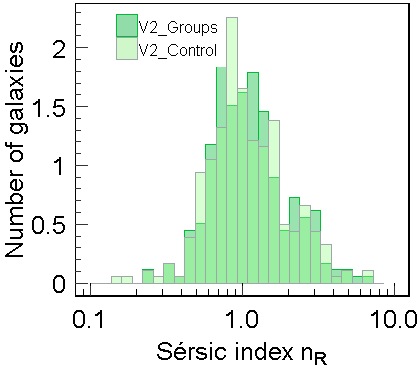}
\includegraphics[width=.66\columnwidth]{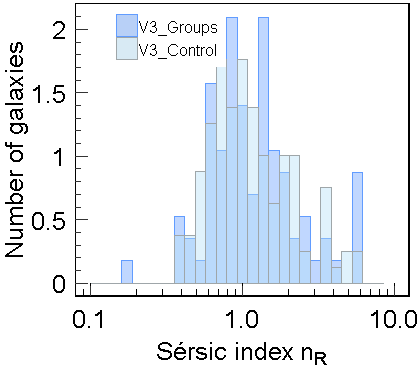}
\caption{From left to right, S\'ersic index histograms for samples V1, V2 and V3. The light and darker colors in each panel correspond to Control and Group samples, respectively. } 
\label{histSersic_ControlGroup_V1V2V3}
\end{figure*}

A decrease in SFR and sSFR in galaxy groups could be originated as a result of a morphology driven quenching process, as suggested by \citet[][]{calvi18}. To test this hypothesis, in Fig.  \ref{histSersic} we show a comparison of the S\'ersic index for group and control galaxies for every one of our three volume limited samples. For the same samples, Table \ref{tab:Sersic} indicates the percentage of galaxies with S\'ersic index  lower and  higher than 2, which is our threshold to divide late from early type morphologies. From Table \ref{tab:Sersic} it is evident that group and control galaxies have very similar percentages. The V3 sample shows the highest percentage of early-type galaxies, however, the percentage of both, group and control in V3 are consistent within 1\%.
Histograms for the S\'ersic index distributions are shown in Fig. \ref{histSersic_ControlGroup_V1V2V3} for control and group galaxies for each volume, in each case following similar distributions. We quantify also the possible differences on the means of the distributions by running a t-test for each volume. For a 95\% confidence level, we cannot reject the null hypothesis of both samples (group and control) having the same distribution. Hence, we discard a morphological driven quenching, since both, control and group galaxies show similar morphologies at given volume. Again, we highlight that this study is focused on Star Forming galaxies with at least four emission lines, and a morphology driven quenching process can still be happening in galaxies that were cut-off form our sample.



\begin{table}
\centering
\caption{Comparison of S\'ersic index for groups and control galaxies. From left to right, Volume sample, total number of galaxies in each volume, percentage of galaxies with S\'ersic index lower than 2 (late-type morphology) and higher than 2 (early-type morphology).}
\begin{tabular}{lrrr}
\hline
\hline
Volume & \# of Galaxies & n$_{\rm R}$ < 2 (\%) & n$_{\rm R}$ > 2 (\%)\\ \hline
V1 Groups/Control & 352/358 & 68.5/68.7 & 14.5/10.9  \\
V2 Groups/Control & 299/322 & 72.0/70.5 & 15.7/12.4  \\
V3 Groups/Control & 105/128 & 65.7/72.6 & 16.2/17.2  \\
\hline
\end{tabular}\label{tab:Sersic}
\end{table}

\section{How much does environment affect the properties of galaxies?}
\label{Environment}



 In this section, we analyse variations in metallicity and star formation rate in new sub-samples created taking into account  the distribution of group members, distance to the group center, $\Sigma_5$, and stellar mass, as explained below.

\begin{figure}
\centering
\includegraphics[width=\columnwidth]{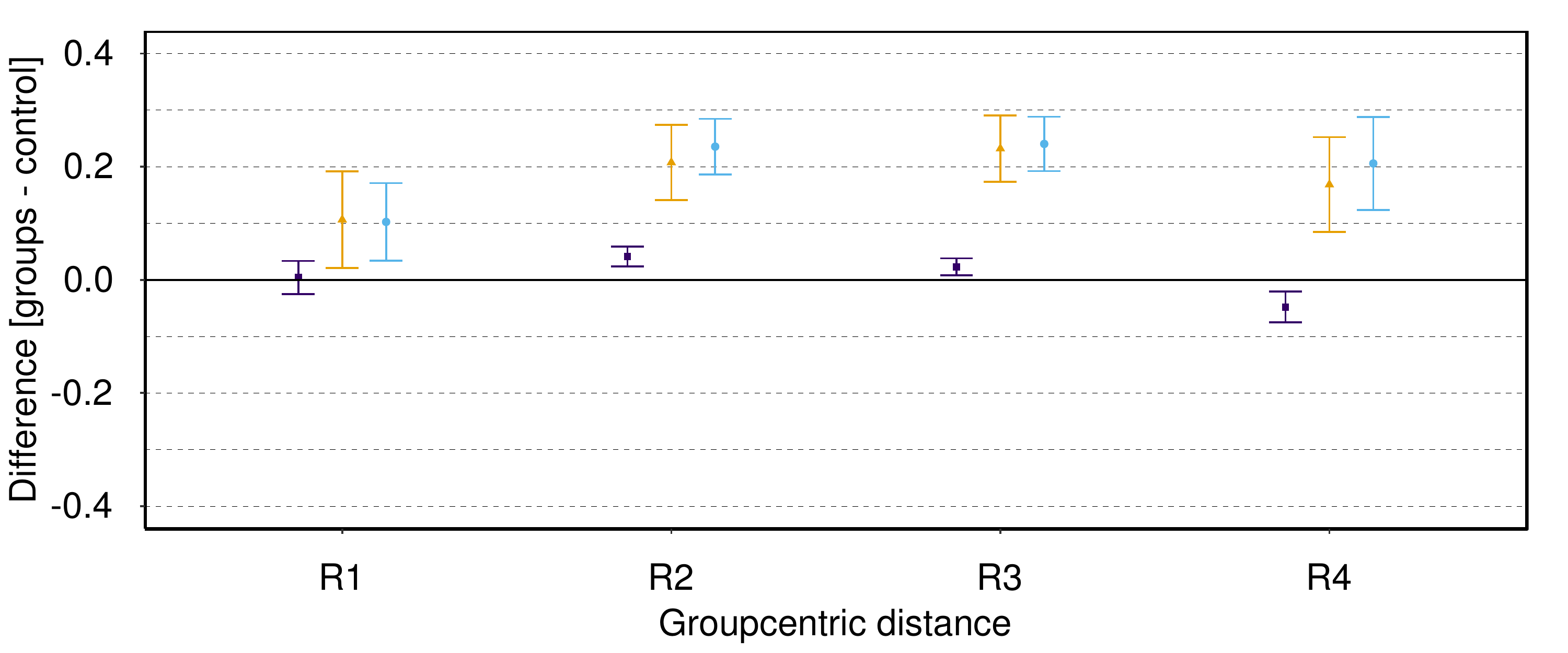}
\caption{Differences for the zero-point coefficients in metallicity, SFR and SSFR (color code as in Figure 9) between groups and control galaxy samples for all three volumes grouped. The median and error bars of one standard deviation are represented for four ranges of group-centric distance from low to high.}
\label{diffDistV123}
\end{figure}

First, to understand the general scope of the data, we combine the three volumes described above into one single sample. We divide the combined sample into four bins of distance to the center of the group (R1 to R4, see Fig. \ref{histDistIterCen}), where R1 ranges from 0.01 to 0.1 Mpc, and R2 to R4 are equally spaced logarithmically, from 0.1 to 1.6 Mpc, as shown in Fig. \ref{diffDistV123}. Our results indicate small variations in gas metallicity for group galaxies with respect to the field, with a maximum difference of $\sim$ -0.05 dex ($\sim 2\sigma$) for galaxies in the most distant bin. In contrast, we always find increments in SFR (and sSFR) with respect to the control sample of field galaxies at all distances, where the smallest increment corresponds to the closest distance bin, and the largest the two intermediate bins. The low metallicity and high SFR in the more distant bin is likely to be related to the presence (or accretion) of \ion{H}{i} rich galaxies residing preferentially in the outskirts of groups  \citep[e.g.,][]{Hess13}.


Other authors find that the SFR is suppressed near the center of groups and clusters \citep[e.g.][]{Poggianti99, Couch01, Barsanti2018}. Our results indicate however, a small increment of 0.1 dex for the most central group galaxies in our sample.
To understand the origin of this discrepancy, it is important to bear in mind that our sample is composed only of SF galaxies, and selected to have at least four emission lines to be able to estimate SFRs and gas metallicities. As already noted, these requirements bias our sample to late-type galaxies (see Fig. \ref{histSersic}), decrease the number of galaxies sampled near group centers, and prevent the inclusion of low-SFR, mostly quenched, and passive galaxies that might otherwise decreas the average SFR of our samples near the group center.
Additionally, some effect may be contributed by the different way of selecting control samples. For instance, some authors compare group with field galaxies without further matching for stellar mass or redshift \citep[e.g., ][]{vulcani10}. Nevertheless, proper control samples are necessary to establish robust differences \citep[e.g.,][]{Ellison09}. 

\begin{figure}
\centering
\includegraphics[width=\columnwidth]{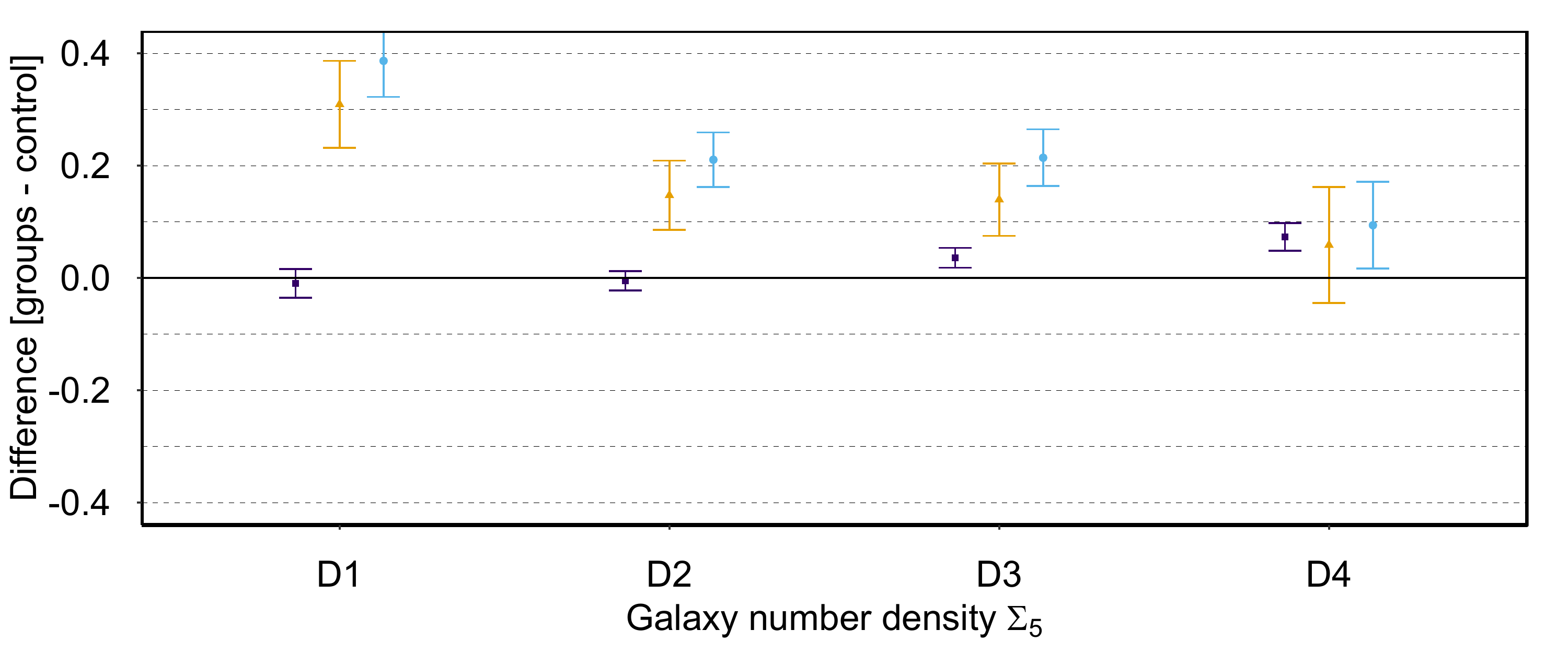}
\caption{Differences for the zero-point coefficients in metallicity, SFR and SSFR (color code as in Figure 9) between groups and control galaxy samples for all three volumes grouped. The median and error bars of one standard deviation are represented for four ranges of $\Sigma_5$ density from low to high.}
\label{diffS5V123}
\end{figure}

To better quantify the influence of local environment, we use the surface number density ($\Sigma_5$) \citep{Muldrew2011}. From the combined sample, we created four bins (D1 to D4) of 0.5 dex covering the total range of values from 0.1 to 100, and estimated the differences in Z, SFR and sSFR, as shown in  Fig. \ref{diffS5V123}. Our data shows enhanced SFRs in group galaxies for all the $\Sigma_5$ bins. The  differences in SFR and sSFR decrease when increasing the surface density, reaching values slightly higher (but still below 0.1 dex) in the bin with the highest $\Sigma_5$.

This supports the accelerated evolution scenario (``in situ evolution") in groups. The differences in SFR and sSFR in the group vs field sample in the highest density and the closest distance bins respectively, indicate that the effect of the environment has mainly occurred as part of the infalling process.

Similarly, the metallicity found in our group sample is higher by $\sim$0.08 dex in the highest surface density bin, in agreement with \citet{Ellison09}, while in the rest of the bins the values are comparable with those in the field. Stripping of low-metallicity gas from the galaxy outskirts, as well as suppression of metal-poor inflows towards the galaxy centre, are key drivers of the enhancement of gas metallicity \citep[e.g.,][]{Bahe2017}.

\begin{figure}
\centering
\includegraphics[width=\columnwidth]{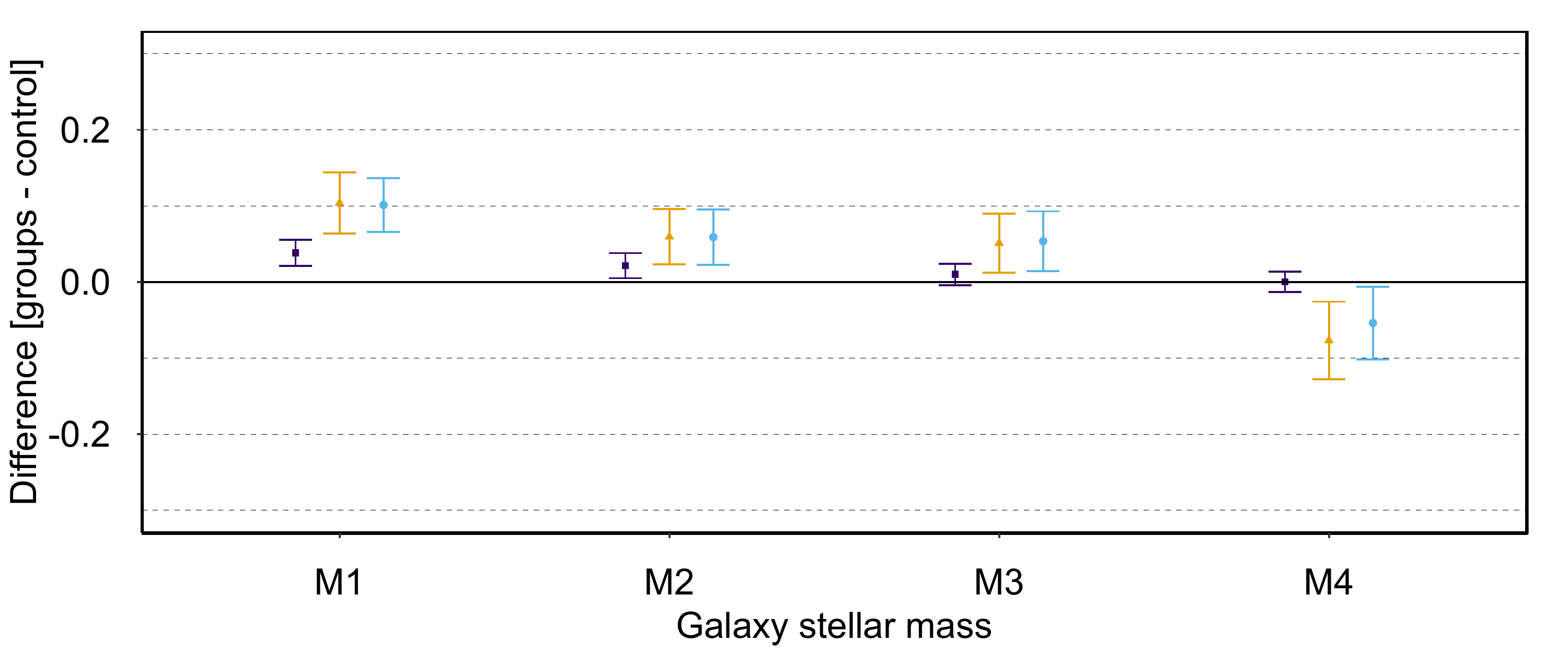}
\caption{Differences for the zero-point coefficients in \mbox{metallicity}, SFR and SSFR (color code as in Figure 9) between groups and control galaxy samples for all three volumes grouped. The median and error bars of one standard deviation are represented for four ranges of stellar mass from low to high.}
\label{diffmassV123}
\end{figure}

Next, we aim to explore the effect of the stellar mass in the joint sample.   
\citet{Moutard18} propose two different types of quenching of the star formation activity. A fast environmental quenching channel followed by young low-mass galaxies, $\log(M_{\star}$/$M_{\odot})$ $<$9.7, and a slow quenching channel followed by more evolved higher-mass galaxies.
In Fig. \ref{diffmassV123} we show the differences in our combined sample divided into four mass bins as indicated in Table \ref{diffVAllV123_ranges_tab}. To accurately test signs of fast quenching, our first mass bin M1 is formed by \mbox{galaxies} with log(M$_{\star}$/$M_{\odot})$ $<$ 9.7, as discussed in \citet{Moutard18}. 
Our results show a small enhancement of $\sim$0.1 dex for SFR and sSFR (see Fig. \ref{diffmassV123}) for the lowest mass bin M1.  If environmental mechanisms are the dominant channel to suppress the star formation in galaxies on a short time-scale \citep{Moutard18}, then we should not expect any significant changes in the star formation compared to the control sample for low mass galaxies. In addition, since such galaxies would quickly be quenched and classified as passive, they would be cut out of our sample. Therefore, we do not observe signs of fast quenching in this mass range for our sample of galaxies. 

For the same low mass bin in Fig. \ref{diffmassV123}, we observe an increment in the gas metallicity by \mbox{$\sim$ 0.05 dex}, while it remains unchanged for the rest of the mas bins. This increment in metallicity has been previously observed in simulations as a signature of ``chemical pre-processing" of infalling cluster galaxies \citep[e.g.,][]{Gupta18}. Under this paradigm, at $z<1.0$, cluster galaxies (both already accreted and infalling) accrete gas that is 2-3 times more metal rich compared to field galaxies. 
Furthermore, since environmental processes are most effective for galaxies with log(M$_{\star}$/$M_{\odot})$ $< 10$  \citep{peng10}, it is likely that inflow of pre-enriched gas drives the observed metallicity enhancement.

\begin{figure}
\centering
\includegraphics[width=\columnwidth]{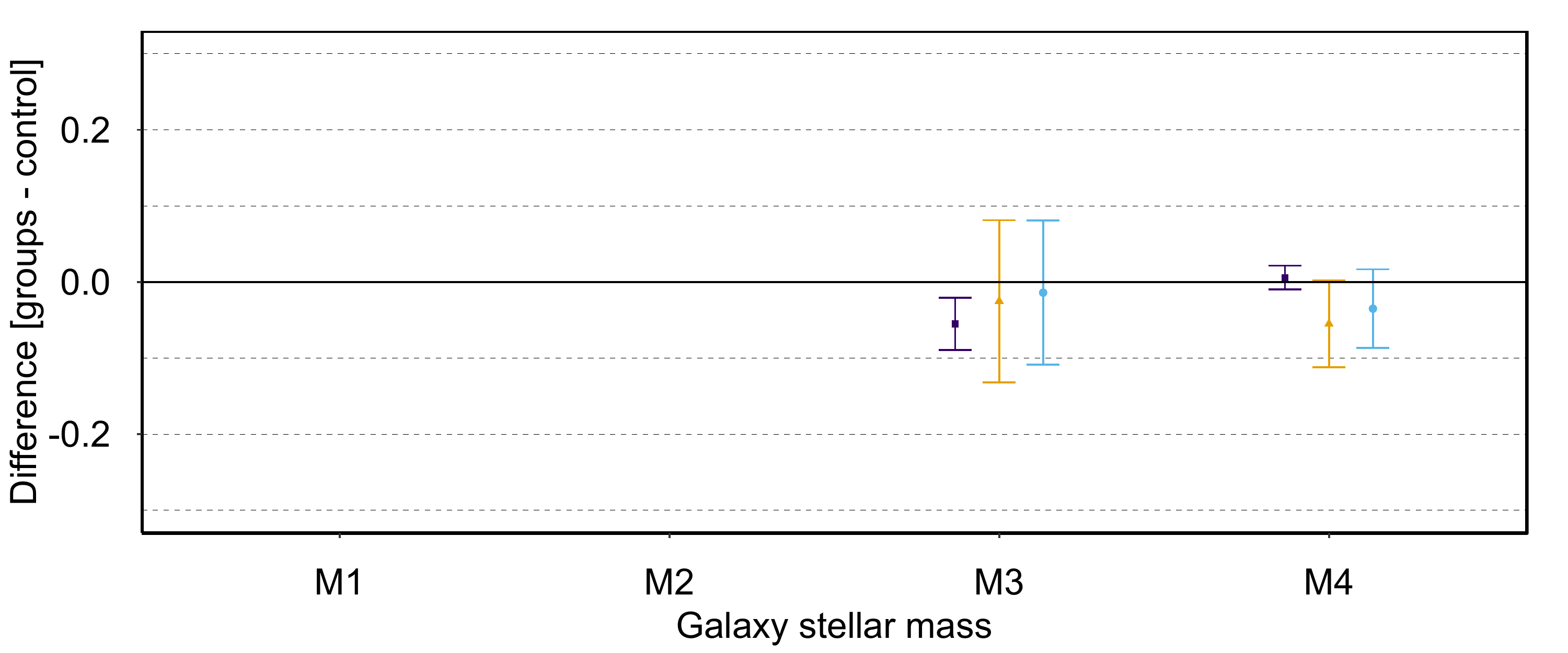}
\caption{Similar to Fig. \ref{diffmassV123}, but using a volume limited sample with the magnitude limits of V3, see text.}
\label{diffmassV123_RangeV3All}
\end{figure}

The SFR and sSFR show only small increments of $\sim$ 0.05 dex for the next mass bins M2 and M3 (see Fig. \ref{diffmassV123}), while for the massive bin M4 there is a clear suppression of SFR and sSFR by $\sim -0.06$ dex. Even though the control samples should ameliorate any selection effect, the combination of all three volume-limited samples together could introduce a bias, since each one of them has a different luminosity and hence stellar mass limit.   
To control for this effect, we created a new volume limited sample by imposing the magnitude limits of V3 on all 3 volume-limited samples. In this way, any luminosity or mass effects should be controlled. Naturally, this new volume-limited sample limits our data to the most massive galaxies, and hence we are only able to analyze samples M3 and M4, the result is shown in Fig. \ref{diffmassV123_RangeV3All}. We are able to recover signs of quenching for the most massive galaxies in M4, and hence corroborate the observed sign of quenching for massive galaxies.

As already mentioned, since our sample of galaxies is restricted to galaxies with at least four emission lines, we are likely missing quenching signatures from more massive and passive galaxies. Therefore, we are unable to address the transition from SF to passive galaxies and hence can not compare our results directly with previous work in that area \citep[e.g., ][]{Wijesinghe12}.





\begin{table}
\caption{Interval ranges of group-centric distance, $\Sigma_5$, and stellar mass for the combination of the three volume limited samples used if Figures \ref{diffDistV123} to \ref{diffmassV123}.}
\begin{tabular}{l r p{1mm} r p{1mm} r p{1mm} r p{1mm} r}
\hline
\hline
 &  & B1 &  & B2 &  & B3 &  & B4 &  \\
Distance & 0.01 & - & 0.10 & - & 0.25 & - & 0.63 & - & 1.60\\
$\Sigma_5$ & 0.41 & - & 2.07 & - & 10.48 & - & 53.12 & - & 269.19 \\
Mass & 8.62 & - & 9.70 & - & 9.99 & - & 10.35 & - & 11.60 \\
\hline
\end{tabular}
\label{diffVAllV123_ranges_tab}
\end{table}

\begin{figure}
\includegraphics[width=\columnwidth]{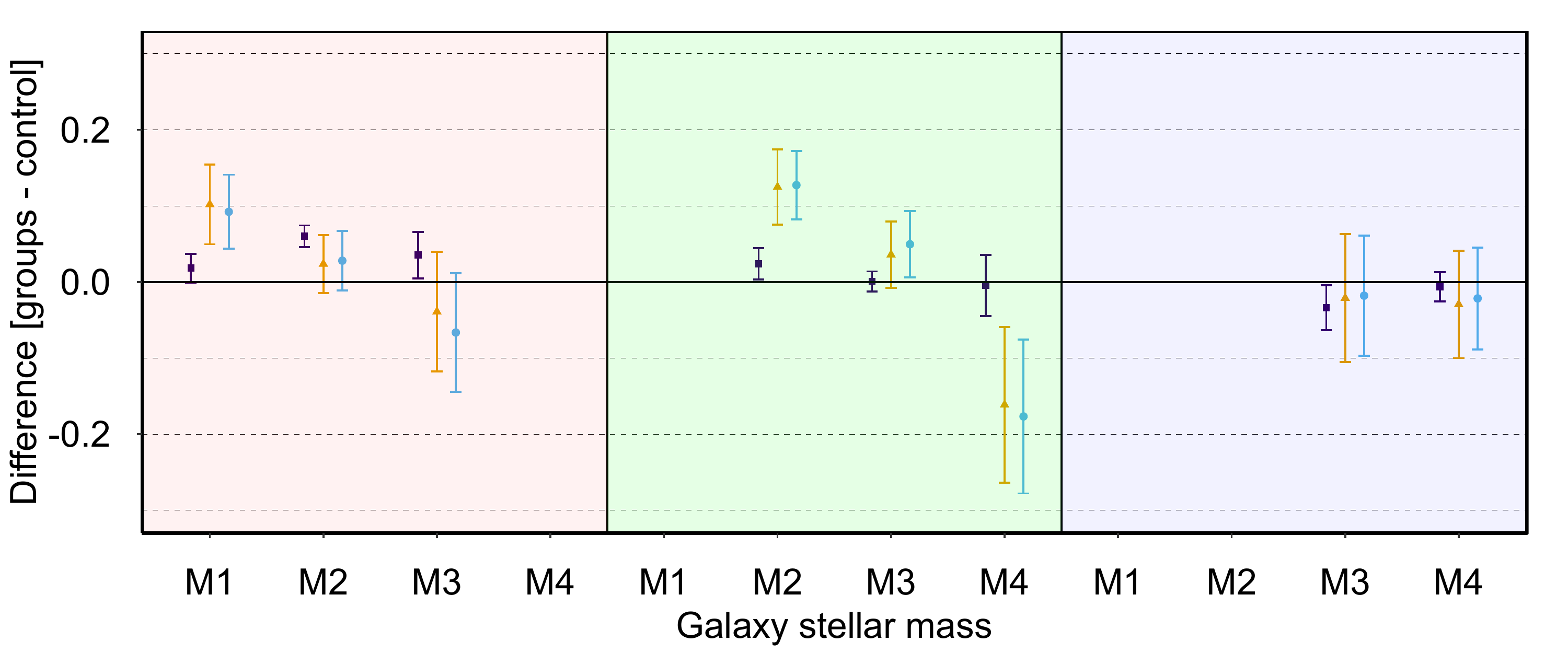}
\caption{Differences for the zero-point coefficients between groups and control galaxy samples, in metallicity, SFR and SSFR (color code as in Figure \ref{diffV}). The median and error bars of one standard deviation are represented. Base volumes and additional sub-samples consisting of the same four bins in mass as in Figure \ref{diffmassV123} (see Table \ref{diffVAllV123_ranges_tab}).}
\label{diffZSFR}
\end{figure}

To observe more detailed signatures of the chemical pre-processing of infalling galaxies or other evolution, we focus now on our individual volume limited samples. Each \mbox{volume} is considered individually, and divided into four equal mass bins. A summary of the mass ranges for each sub-sample is given in Table \ref{diffVsamples_tab}, and the differences found in \mbox{Fig. \ref{diffZSFR}}. We find consistent enhancements in gas metallicity for the volume V1. This result is in agreement with the ``chemical pre-processing" scenario described above, where the effect of environment is stronger in low mass galaxies.

On the other hand, at all volumes, the SFR shows signs of quenching for the most massive galaxies in the bin M4 of V3. The mass range M3 do not show any clear pattern, showing a quenched SFR for volumes V1 and V3, and an enhanced SFR for V2. We do not discard however, an evolutionary effect between volumes V1 and V2, as suggested by the common mass ranges M2 and M3. These mass ranges show higher metallicities, and lower SFRs in V1, in contrast with negligibly changes in metallicity, and higher SFRs for the same mass ranges in V2.

It is likely that the chemical enrichment of the intra-cluster medium (ICM) plays a major role in enhancing the gas metallicities of infalling galaxies, as observed in simulations \citep[e.g.,][]{Gupta18}. On the other hand, there is evidence that the metallicity of the ICM does not evolve up to redshift $z \sim 1.0$ \citep{McDonald16, Biffi17}.

\begin{table*}
\centering
\caption{Summary of the differences found for the different sub-samples. From left to right, the columns indicate the sub-sample, number of galaxies, median S\'ersic index \~n$_{\rm r}$, surface number density, median stellar mass, 95 $\%$ highest density interval (HDI), and difference in gas metallicity (Z), SFR and sSFR, respectively. The first, second and third block correspond to the V1, V2 and V3 volumes, respectively.}
\begin{tabular}{lrrrrcrrr}
\hline
\hline
sub-sample  & No. of galaxies &  S\'ersic \~n$_{\rm r}$  & $\Sigma_5$ & $\mathrm{ \tilde{M}_{\odot}}$ &  95\% HDI ($\mathrm{M_{\odot}}$) & $\Delta$Z & $\Delta$SFR & $\Delta$sSFR\\ \hline

V1 & 352 & 1.12 & 14.63 & 9.71 & 9.07 - 10.57 & 0.040 $\pm$ 0.011 & 0.009 $\pm$  0.036 & 0.030 $\pm$  0.031\\

M1 & 94 & 1.09 & 15.15 &  9.34 & 9.13 - 9.46 & 0.018 $\pm$ 0.019 & 0.102 $\pm$ 0.052 & 0.092 $\pm$ 0.049\\
M2 & 180 & 1.10 & 14.08 &  9.73 & 9.48 - 9.71 & 0.060 $\pm$ 0.014 & 0.024 $\pm$ 0.038 & 0.028 $\pm$ 0.039\\
M3 & 55 & 1.25 & 13.87 &  10.19 & 9.72 - 9.94 & 0.035 $\pm$ 0.031 & -0.039 $\pm$ 0.079 & -0.066 $\pm$ 0.078\\
M4 & 17 & 1.60 & 18.62 &  - & - & - & - & -\\
\hline

V2 & 299 & 1.05 & 6.81 &  10.06 & 9.59 - 10.76 & 0.005 $\pm$ 0.013 & 0.043 $\pm$ 0.045 & 0.057 $\pm$ 0.032\\

M1 & 4 & 3.18 & 5.66 &  - & - & - & - & - \\
M2 & 107 & 0.93 & 7.04 &  9.84 & 9.48 - 9.71 & 0.024 $\pm$ 0.021 & 0.125 $\pm$ 0.050 & 0.127 $\pm$ 0.045\\
M3 & 144 & 1.12 & 6.60 &  10.18 & 9.72 - 9.94 & 0.001 $\pm$ 0.013 & 0.036 $\pm$ 0.044 & 0.050 $\pm$ 0.043\\
M4 & 37 & 1.39 & 8.91 &  10.66 & 9.95 - 10.82 & -0.004 $\pm$ 0.040 & -0.161 $\pm$ 0.102 & -0.177 $\pm$ 0.101\\

\hline

V3 & 105 & 1.06 & 2.32 &  10.64 & 10.17 - 11.11  & -0.024 $\pm$ 0.017 & -0.060 $\pm$ 0.068 & 0.032 $\pm$ 0.051\\
M1 & 0 & - & - &  - & - & - & - & -\\
M2 & 2 & 0.55 & 2.33 &  - & - & - & - & -\\
M3 & 33 & 0.85 & 1.82 &  10.37 & 9.72 - 9.94 & -0.034 $\pm$ 0.030 & -0.021 $\pm$ 0.084 & -0.018 $\pm$ 0.079\\
M4 & 63 & 1.29 & 2.23 &  10.75 & 9.95 - 10.82 & -0.006 $\pm$ 0.019 & -0.029 $\pm$ 0.071 & -0.022 $\pm$ 0.067\\

\hline
\end{tabular}
\label{diffVsamples_tab}
\end{table*}

\begin{table*}
\centering
\caption{Summary of the differences found for the different sub-samples. Sub-samples correspond to the total sample of group galaxies (V1+V2+V3). In each of the four blocks, sub-samples are selected as follows: group-centric distance, surface number density, galaxy stellar mass, and galaxy stellar mass with a magnitude-cut corresponding to V3. From left to right, the columns indicate the sub-sample, number of galaxies, median S\'ersic index \~n$_{\rm r}$, median and 95 $\%$ highest density interval (HDI) of the parameter in consideration, and differences in gas metallicity (Z), SFR and sSFR, respectively.}
\begin{tabular}{lrrrrcrrr}
\hline
\hline
sub-sample  & No. of galaxies &  S\'ersic \~n$_{\rm r}$  & $\mathrm{ \tilde{Dist}}$ &  95\% HDI ($\mathrm{Dist}$) & $\Delta$Z & $\Delta$SFR & $\Delta$sSFR\\ \hline


R1 & 84 & 1.16  &  0.062 & 0.019 - 0.099 & 0.004 $\pm$ 0.029 & 0.106 $\pm$ 0.085 & 0.102 $\pm$ 0.069\\
R2 & 259 & 1.01  &  0.176 & 0.104 - 0.248 & 0.041 $\pm$ 0.018 & 0.207 $\pm$ 0.066 & 0.235 $\pm$ 0.049\\
R3 & 321 & 1.14  &  0.381 & 0.252 - 0.586 & 0.023 $\pm$ 0.015 & 0.2329 $\pm$ 0.058 & 0.240 $\pm$ 0.048\\
R4 & 63 & 0.94  & 0.820 & 0.638 - 1.438 & -0.048 $\pm$ 0.027 & 0.168 $\pm$ 0.084 & 0.206 $\pm$ 0.082 \\

\hline

sub-sample  & No. of galaxies &  S\'ersic \~n$_{\rm r}$  & $\Sigma_5$ &  95\% HDI ($\mathrm{\Sigma_5}$) & $\Delta$Z & $\Delta$SFR & $\Delta$sSFR\\ \hline


D1 & 112 & 0.96  &  2.176 & 1.094 - 3.008 & -0.010 $\pm$ 0.025 & 0.309 $\pm$ 0.077 & 0.386 $\pm$ 0.064 \\
D2 & 276 & 1.20  &  6.135 & 3.172 - 9.529 & -0.005 $\pm$ 0.017 & 0.147 $\pm$ 0.061 & 0.210 $\pm$ 0.049 \\
D3 & 261 & 1.02  &  16.129 & 10.047 - 27.822 & 0.036 $\pm$ 0.018 & 0.139 $\pm$ 0.064 & 0.214 $\pm$ 0.051 \\
D4 & 82 & 1.13  &  43.214 & 32.043 - 88.228 & 0.073 $\pm$ 0.025 & 0.059 $\pm$ 0.103 & 0.094 $\pm$ 0.077 \\

\hline
sub-sample  & No. of galaxies &  S\'ersic \~n$_{\rm r}$  & $\mathrm{\tilde{M}_{\odot}}$ &  95\% HDI ($\mathrm{M_{\odot}}$) & $\Delta$Z & $\Delta$SFR & $\Delta$sSFR\\ \hline


M1 & 189 & 1.06 & 9.473 & 9.136 - 9.690 & 0.038 $\pm$ 0.017 & 0.104 $\pm$ 0.040 & 0.101 $\pm$ 0.035 \\
M2 & 189 & 0.99 &  9.834 & 9.702 - 9.972 & 0.021 $\pm$ 0.016 & 0.059 $\pm$ 0.037 & 0.059 $\pm$ 0.036 \\
M3 & 189 & 1.36 &  10.130 & 9.995 - 10.337 & 0.010 $\pm$ 0.014 & 0.051 $\pm$ 0.039 & 0.054 $\pm$ 0.039\\
M4 & 189 & 1.46 &  10.633 & 10.353 - 11.053 & 0.000 $\pm$ 0.013 & -0.077 $\pm$ 0.051 & -0.054 $\pm$ 0.048\\

\hline
sub-sample  & No. of galaxies &  S\'ersic \~n$_{\rm r}$  & $\mathrm{\tilde{M}_{\odot}}$ &  95\% HDI ($\mathrm{M_{\odot}}$) & $\Delta$Z & $\Delta$SFR & $\Delta$sSFR\\ \hline


M1 & - & - & - & - & - & - & - \\
M2 & - & - & - & - & - & - & -  \\
M3 & 21 & 0.80  &  10.24 & 9.999 - 10.333 & -0.055   $\pm$  0.034 & -0.025 $\pm$ 0.106 & -0.014 $\pm$ 0.095 \\
M4 & 111 & 1.24 &  10.685 & 10.360 - 11.055 & 0.006 $\pm$ 0.016 & -0.055 $\pm$ 0.057 & -0.035 $\pm$ 0.052 \\

\hline
\end{tabular}
\label{tab:diffVAllsamples}
\end{table*}

\section[]{Simulations}\label{Simulations}


We perform a similar study using the IllustrisTNG cosmological magneto-hydrodynamical simulation of galaxy formation \citep{Naiman2018,Nelson2018,Marinacci2018,Pillepich2018,Springel2018}. Following our observational analysis, we constructed similar samples in redshift and absolute magnitude to analyse variations in metallicity and SFR against stellar mass. For  simulated 
cubic volumes of roughly 300 comoving Mpc side length, and starting with a redshift value of 127, a total of 100 snapshots (including all the information for all particles in the whole volume) of the complete temporal evolution are stored. They are separated by time steps ranging from 50 to $\approx$ 100 Ma. The identification of halos and subhalos in groups is performed with a friends-of-friends algorithm. This is performed only on the dark matter (DM) particles. The other types of particles present in the simulation, which include gas, stars or black holes, join the same group of their nearest DM particle.

We take the following approach to select the samples from the simulation. We choose the snapshot whose redshift value is closest to the median value of each of the three volume limited samples from GAMA (Table \ref{tab:Vsimul}). Since in GAMA we have an observational limit imposed by the instrumentation ($r < 19.8$), we have to select, for each snapshot, only the galaxies that could have been observed at their respective redshift value. Limits of absolute magnitude for the r-band are shown also in Table \ref{tab:Vsimul}.

The resolution of the simulation will impose one limit on the galaxies that can be selected for our sample. Studies where morphological characteristics are important define a lower bound of $\approx$ 1000 \citep{Yun2019} or 10000 \citep{Semczuk2020} star particles. Since we are exploring integrated properties we can set a lower limit of 10 particles, which allows us to reach a stellar mass limit in the selected galaxies as low as 10$^9$ M$_{\odot}$. This is important to enable us to analyse the less massive galaxies present in V1. \citet{patton2020} define a lower bound of 90 star particles per galaxy in their recent studies with the IllustrisTNG300 simulation, a similar order of magnitude to ours.

\begin{figure*}
\includegraphics[scale=0.38]{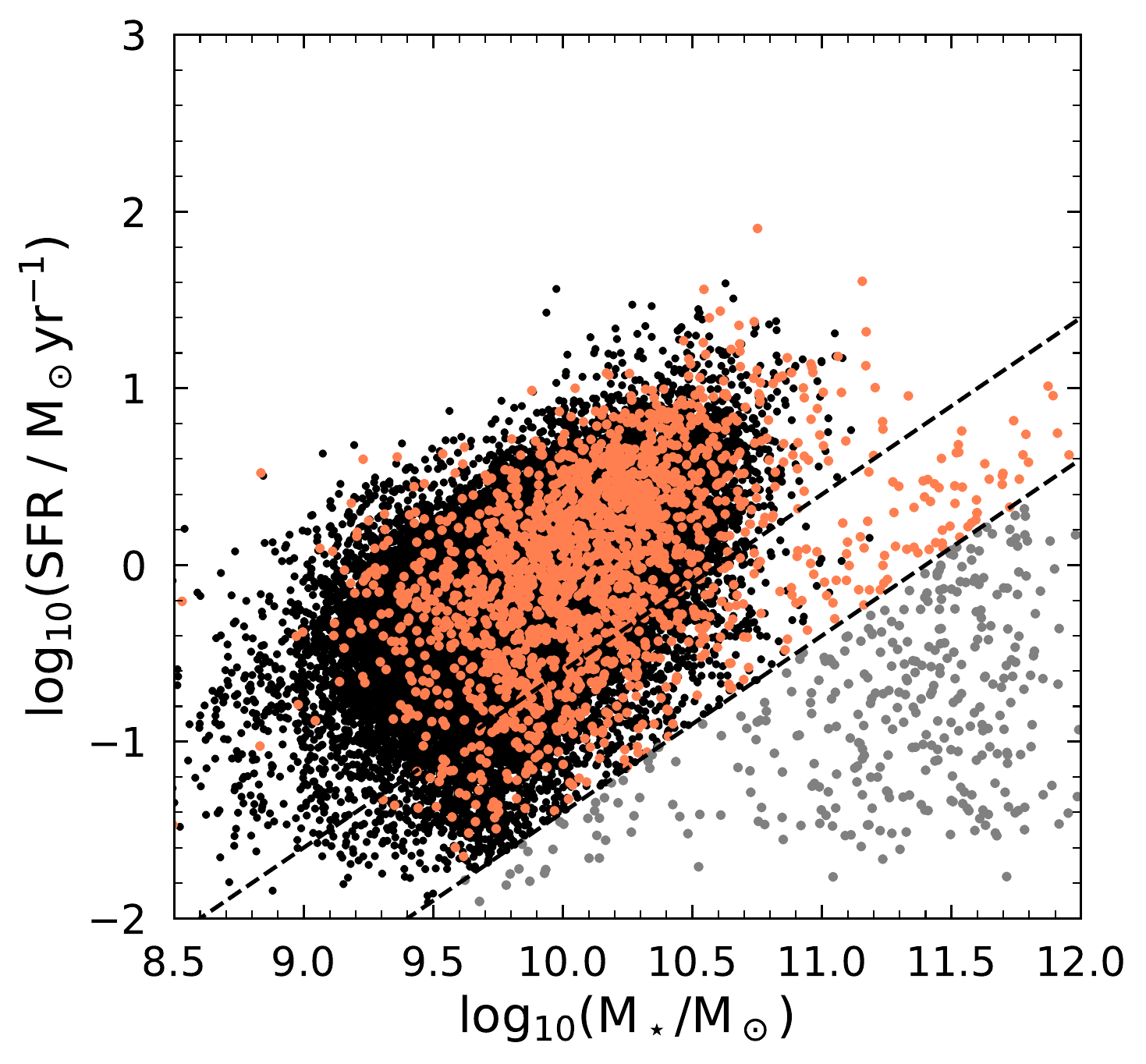}
\includegraphics[scale=0.38]{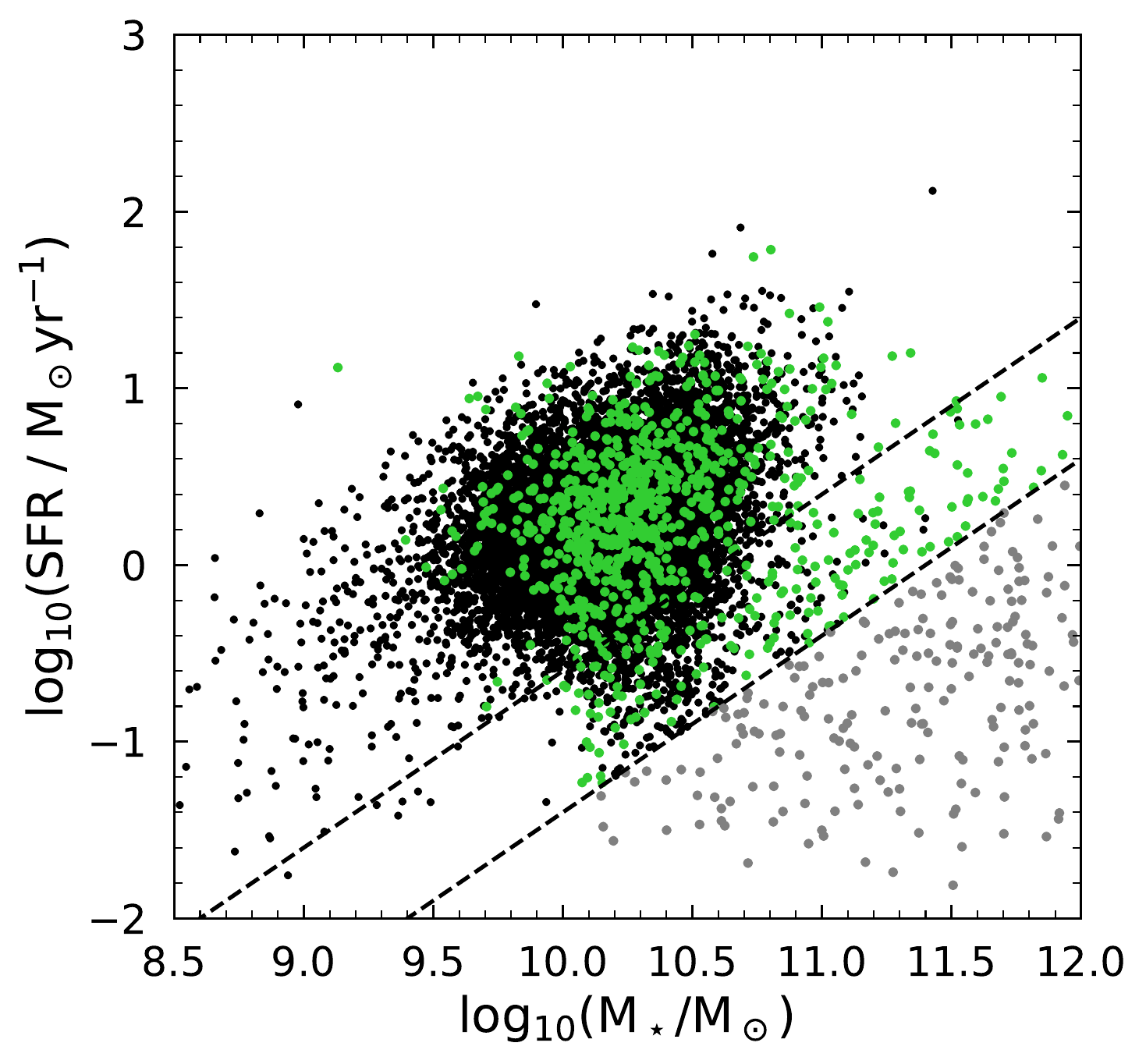}
\includegraphics[scale=0.38]{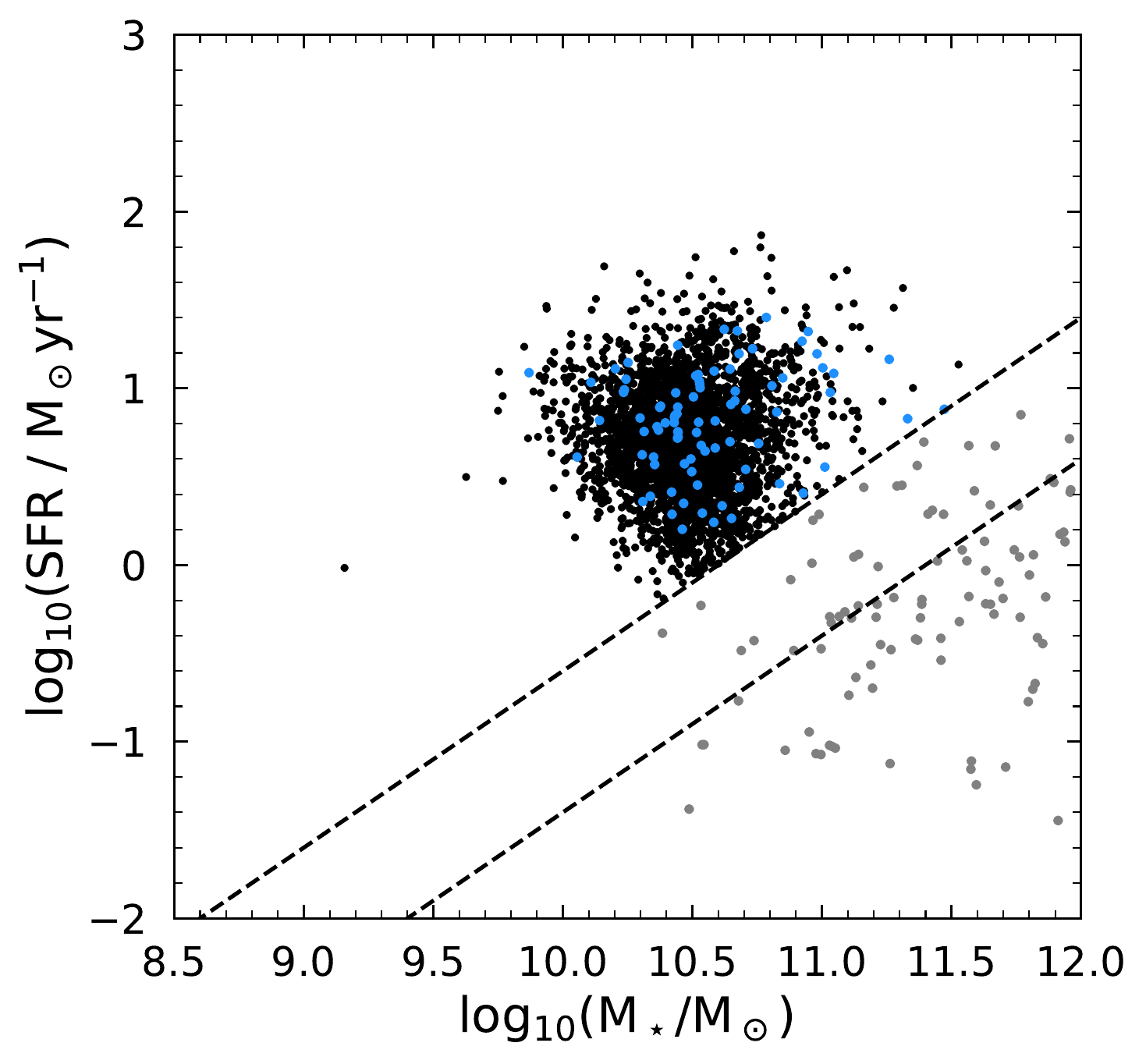}
\caption{M-SFR relationship for our three volume samples V1, V2, V3 (left, centre, right, respectively) from IllustrisTNG300-2.  For the star-forming sample (coloured for groups, black for field), we perform two different selections to reject passive galaxies: ($i$) all galaxies with log(sSFR) higher than -10.6  for V1 and V2, and ($ii$)  all galaxies with log(sSFR) higher than -11.4  for V3. Galaxies considered passive are colored as grey dots.}
\label{MSFR_V123_simul}
\end{figure*}

Similar to the GAMA samples, we want to identify groups with at least 6 members. The galaxy counting for each snapshot is performed before applying the same magnitude limits of the GAMA volumes  (Table \ref{Vsamples_tab}).  
Additionally, we remove passive galaxies using the limit provided by \citet{Hsieh2017} in the M-SFR relation. We can apply two cuts with constant sSFR (depicted as black dashed lines in the M-SFR relation in Fig. \ref{MSFR_V123_simul}). Next we perform two different selections. For the V1 and V2 samples, galaxies with log(sSFR)$>-10.6$  are selected, while galaxies with log(sSFR)$<-11.4$ are considered passive galaxies. For the V3 volume, all galaxies with log(sSFR)$<-10.6$ are considered passive. This first selection provides a more complete but also more contaminated sample for V1 and V2. For the V3 volume, we get a complete and low contamination sample.
Since the simulation does not allow for selecting star-forming galaxies using the BPT diagram, we have additionally matched, for each galaxy in the GAMA group samples, the closest galaxy in mass for the corresponding group samples in the simulation.

The creation of  control samples requires the identification of field galaxies. Since the identification of groups in IllustrisTNG is performed with a FoF algorithm (as is also the case for the GAMA Groups Catalogue, see Section \ref{SampleSelection}), all galaxies in unitary groups are included in the field catalogue. Moreover, all satellite galaxies in halos at distances from the host galaxy larger than 4.5 $\times$ R$_{200c}$ (where R$_{200c}$ is the radius of the sphere with a density of 200 times the critical density of the universe), are also considered to be field galaxies, as suggested by \citet{Busha2015}, \citet{Haines2015} and \citet{Barsanti2018}.

This last step completely defines the group and field catalogues from the simulations. To define the control samples we follow the same procedure as for the observational data from GAMA. We find 3 matches in mass (redshift will be exactly the same for each snapshot) from the field catalogue until the mass distribution is the
same for groups and field galaxies.

\begin{figure*}
\includegraphics[scale=0.34]{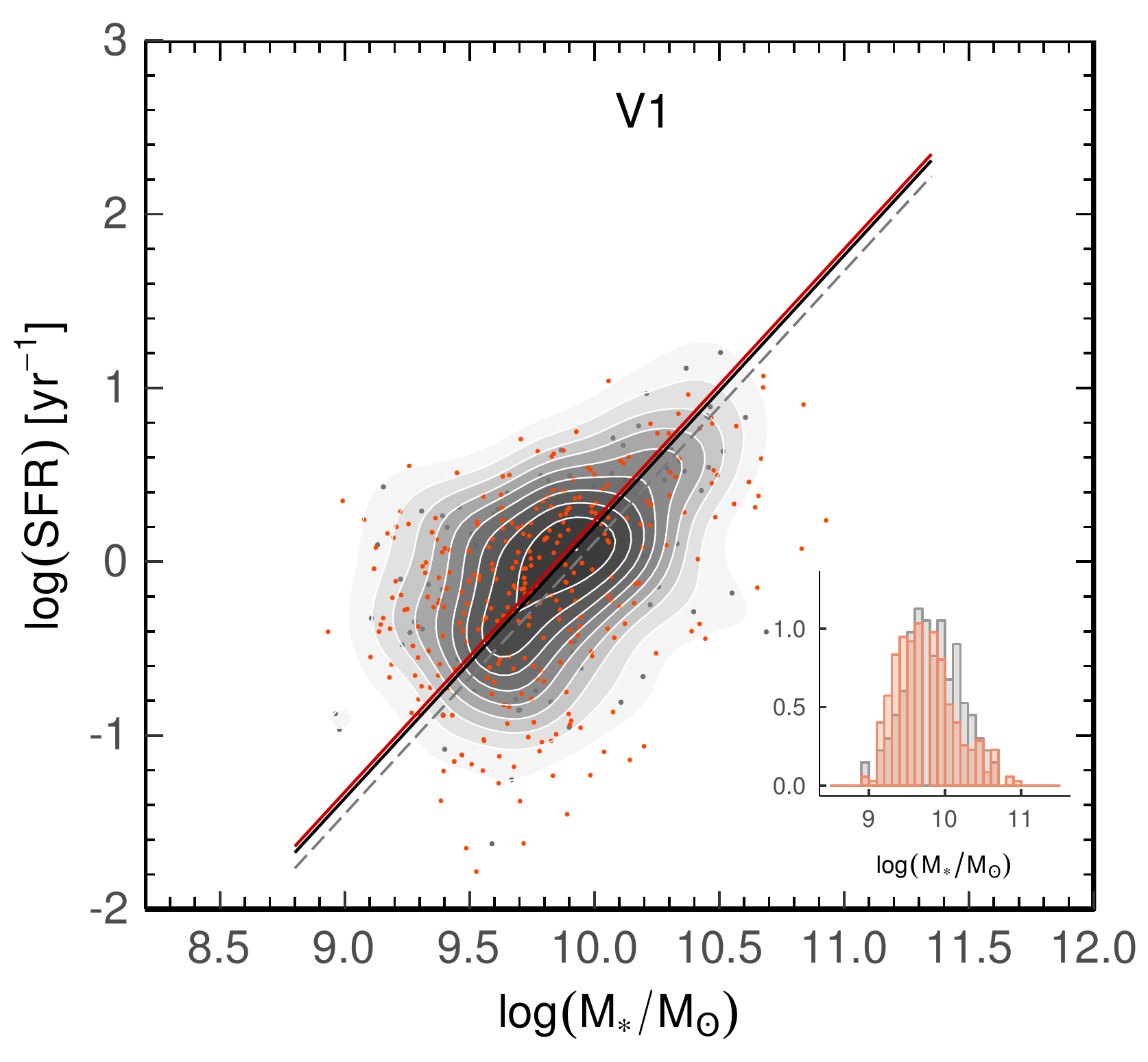}
\includegraphics[scale=0.34]{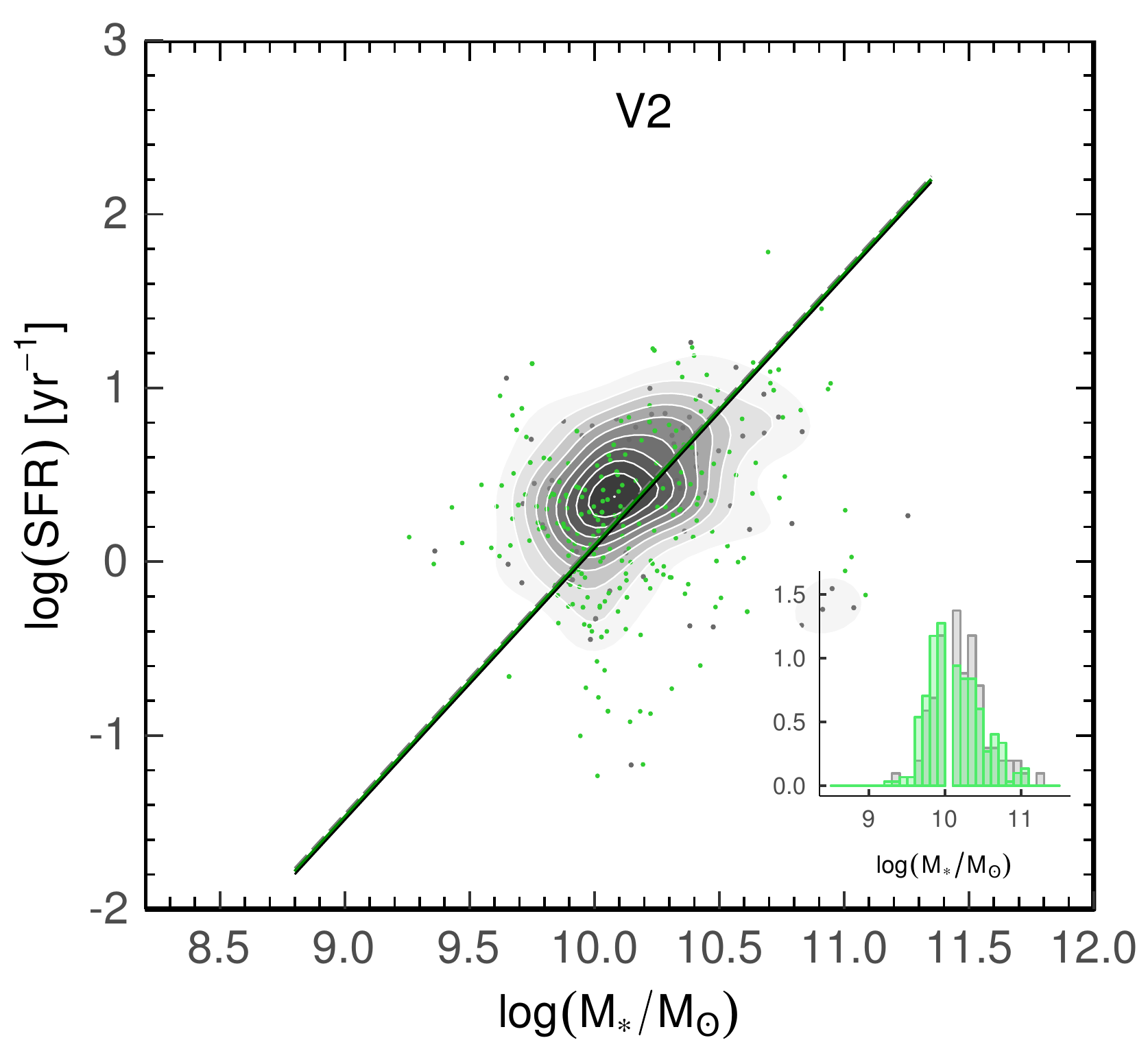}
\includegraphics[scale=0.34]{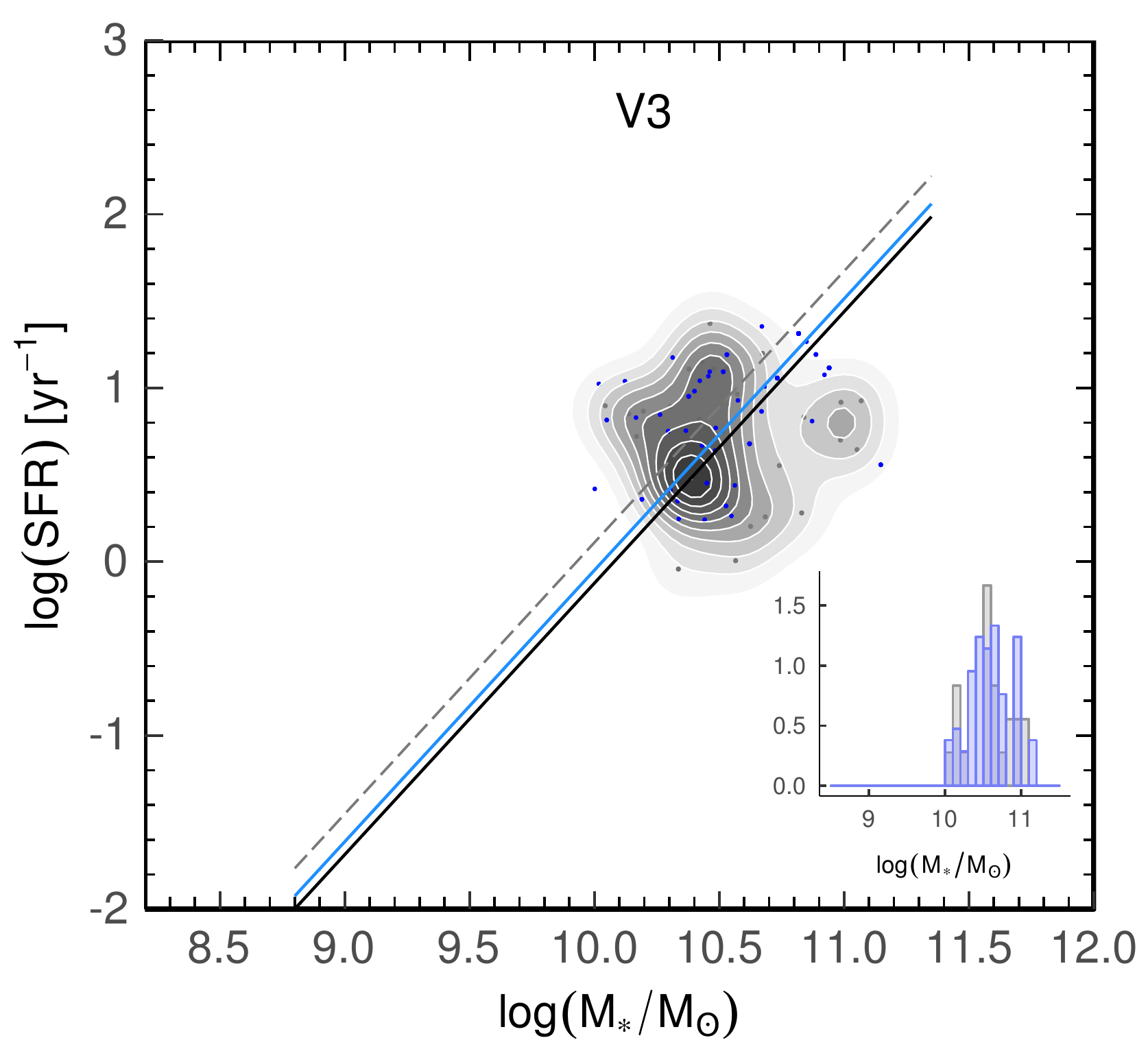}
\caption{M-SFR relationship for the three volume limited samples from simulations (see text) V1, V2, V3 (left, centre, right, respectively). Grey dots represent control group galaxies, and coloured dots (red, green, blue for V1, V2, V3, respectively), group galaxies. The black line represents the best fit for the control sub-sample whereas the coloured line is the fit for the group galaxies sub-sample. The gray dashed line is the fiducial fit to the joint set of the control samples. Inset histograms represent   the mass distributions of the groups (coloured) and control (grey) galaxies.}
\label{_MSFR_V123_simul}
\end{figure*}

\begin{figure*}
\includegraphics[scale=0.34]{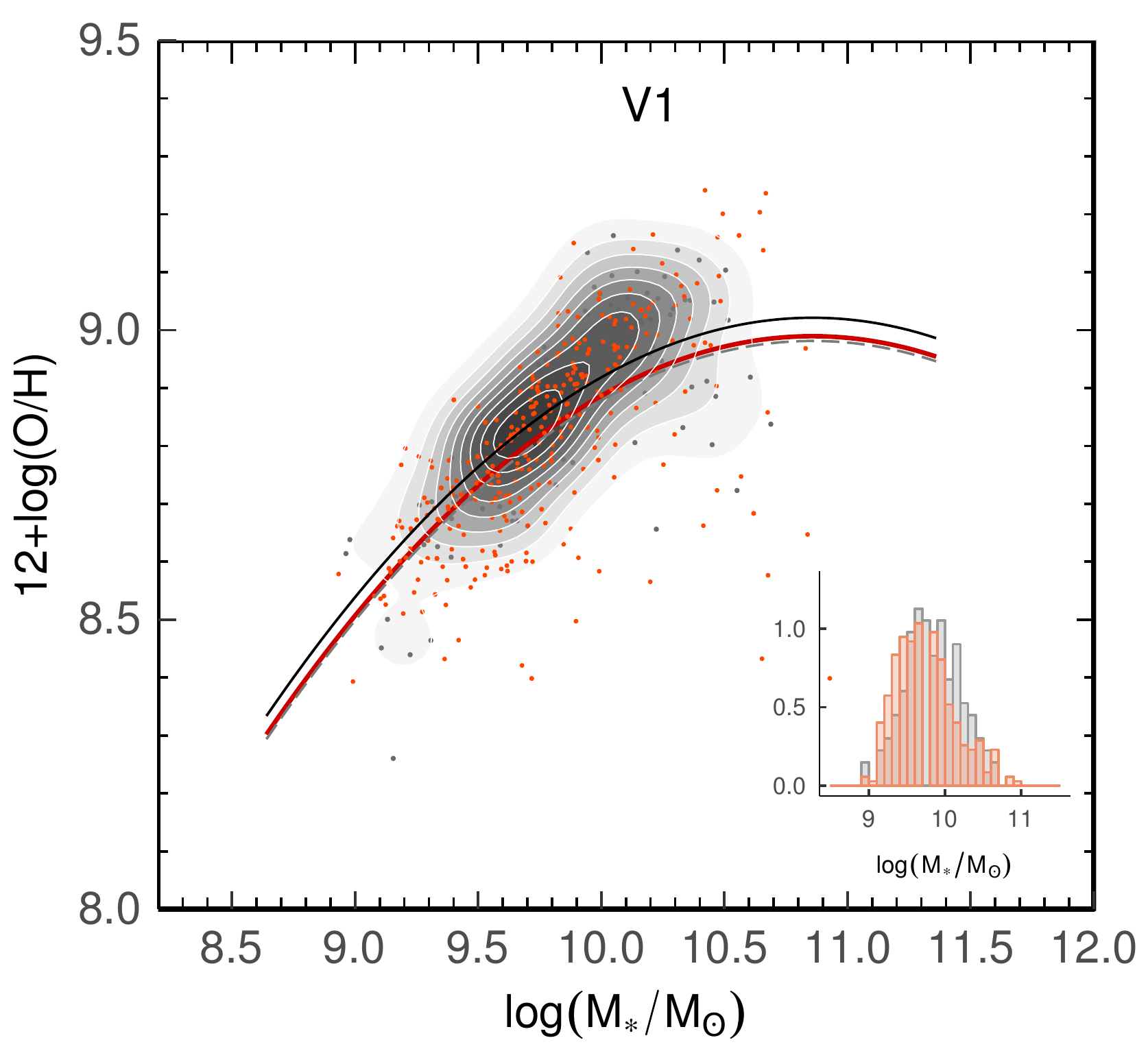}
\includegraphics[scale=0.34]{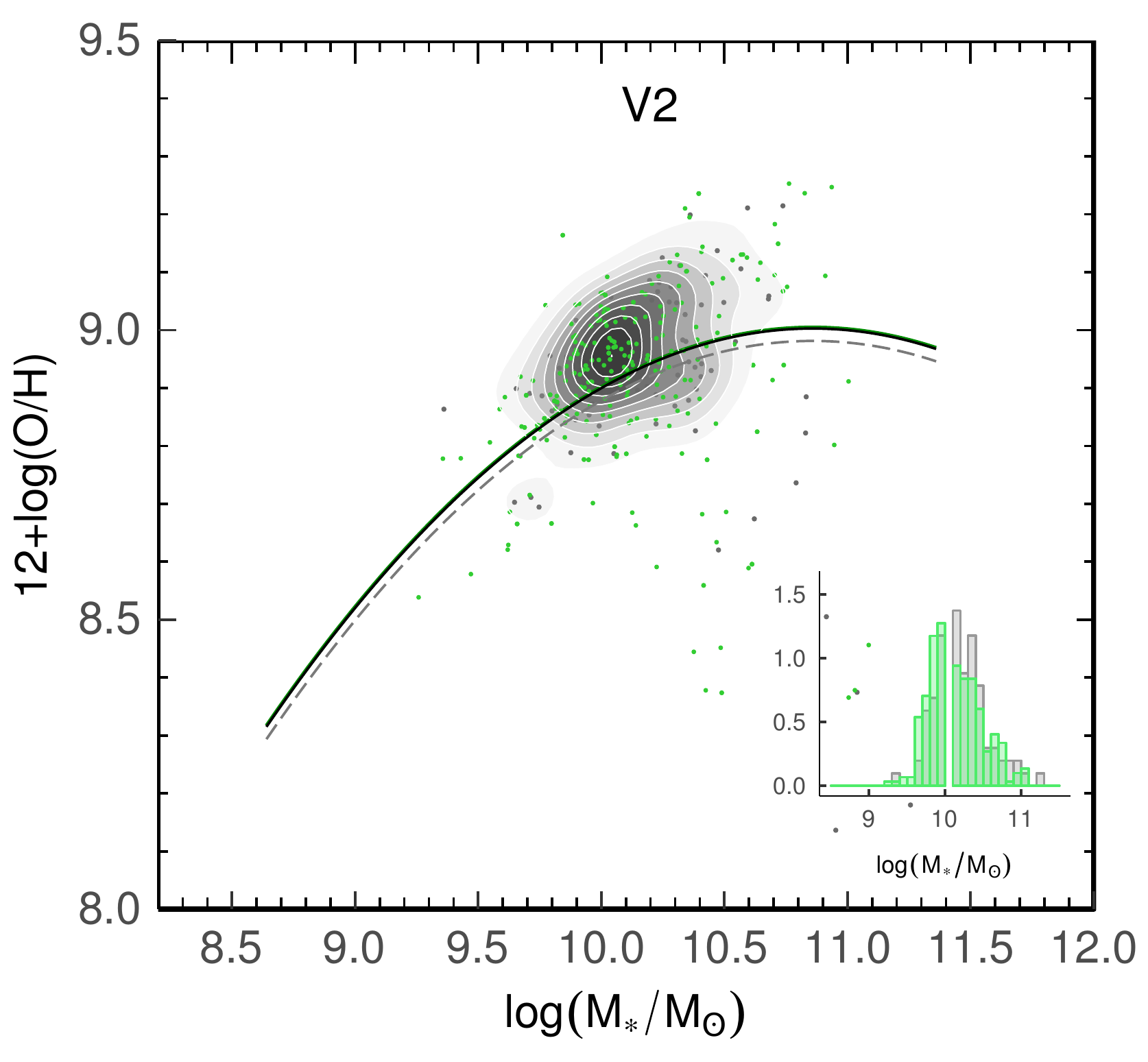}
\includegraphics[scale=0.34]{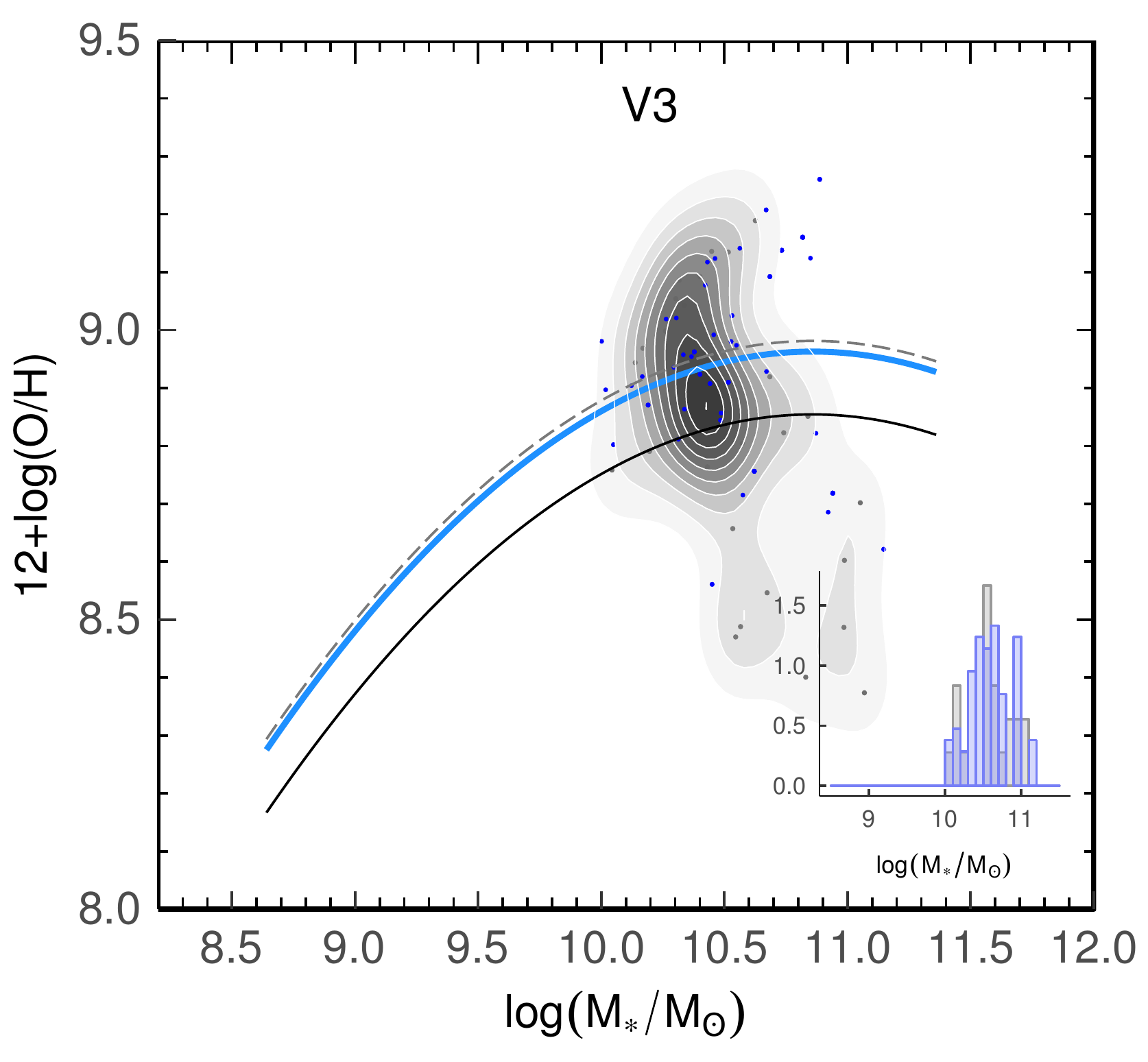}
\caption{M-Z relations for the three volumes from simulations. Symbols and lines similar to Figure \ref{_MSFR_V123_simul}}
\label{_MZ_V123_simul}
\end{figure*}

\begin{figure}
\centering
\includegraphics[width=\columnwidth]{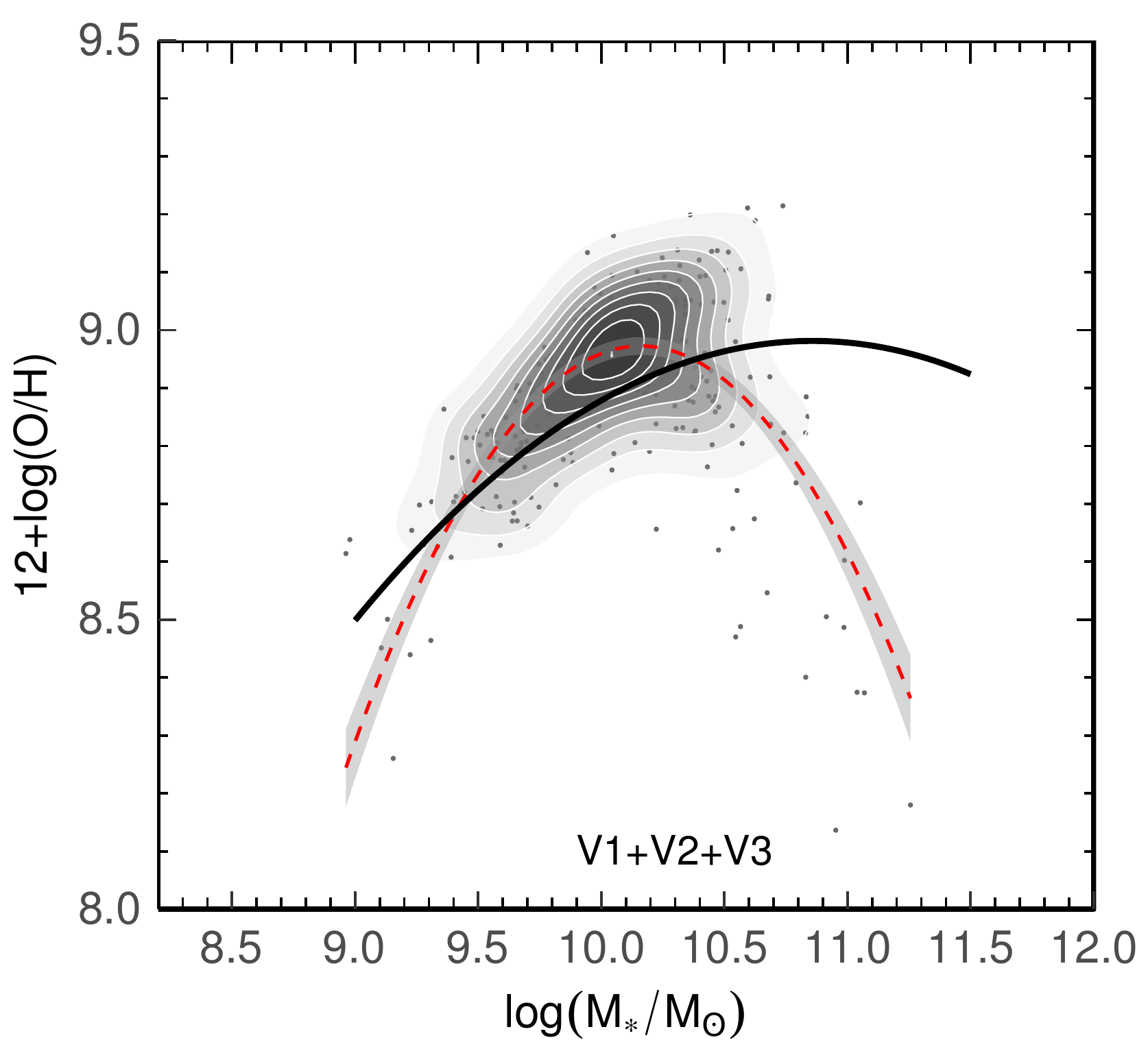}
\caption{M-Z relation for control galaxies from IlustrisTNG. The grey data represent data from the V1, V2 and V3 volumes together. The red line is the best-fitting to the simulated data. The black line is the best-fitting based on the GAMA data (also shown as gray dashed lines in Fig. \ref{_MZ_V123_simul}).} \label{MZAll_simul}
\end{figure}

\begin{figure}
\centering
\includegraphics[width=\columnwidth]{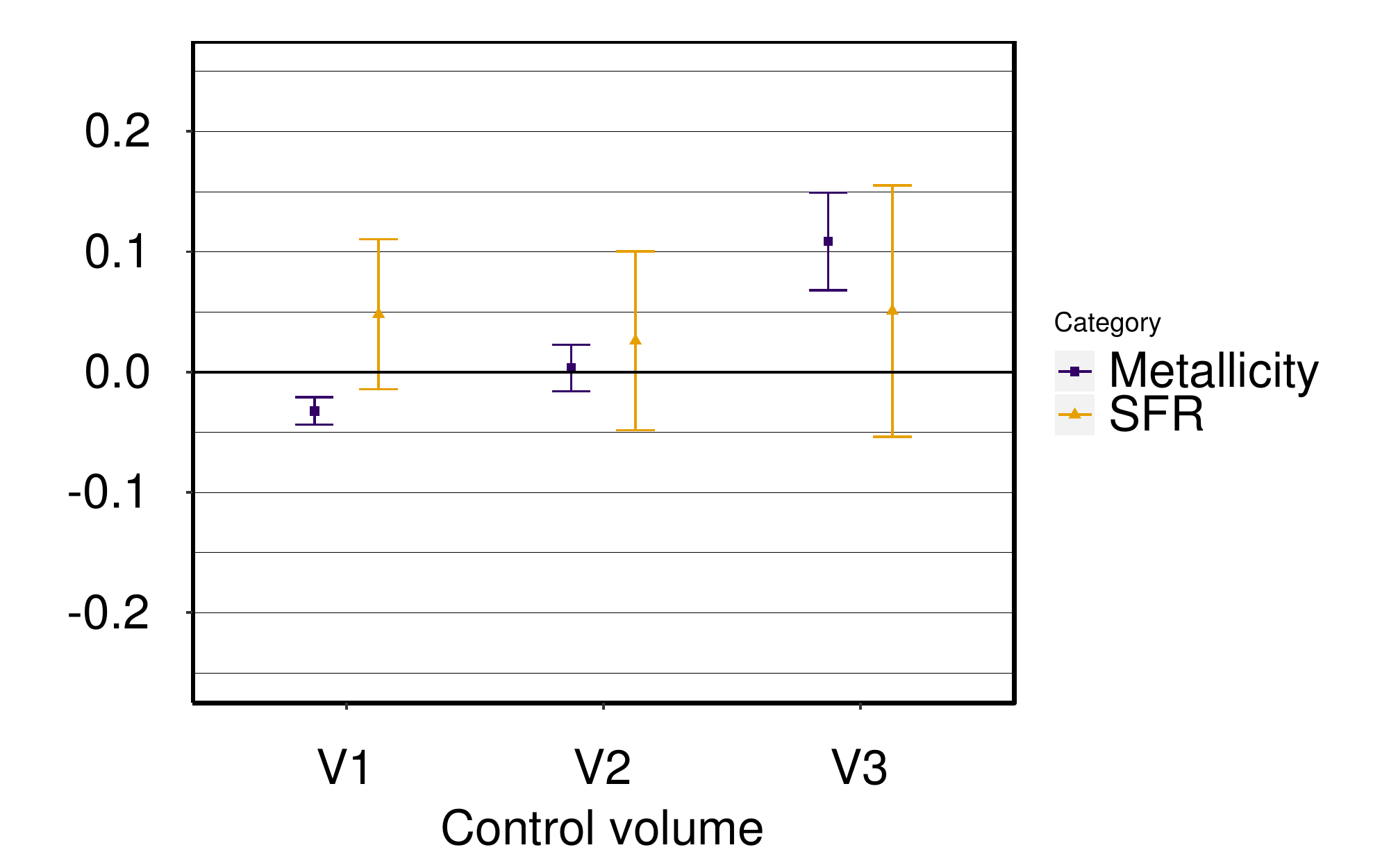}
\caption{Differences for the zero-point between group and control galaxies in the simulation TNG300. The differences and 1-$\sigma$ error bars in metallicity and SFR are color coded as in Fig. \ref{diffV}.} \label{diffV_simul}
\end{figure}

To calculate the M-Z and M-SFR relations we proceed as described in sections \ref{MZgama} and \ref{MSFRgama}. The resulting M-SFR relationships for the different volumes are shown in \mbox{Fig. \ref{_MSFR_V123_simul}}, and in \mbox{Fig. \ref{_MZ_V123_simul}}  for the M-Z relation. For the fit of the M-Z relation, the polynomial corresponding to the GAMA control sample has been taken, leaving free the constant term to match the simulation sample. The GAMA fit was used since the fit to the IllustrisTNG control galaxies is strongly affected by high mass and low metallicity galaxies (red fit in Fig. \ref{MZAll_simul}), that are not present in the GAMA samples.
The differences in Z and SFR  are summarised in the Fig. \ref{diffV_simul}, represented on the same scale as Figure \ref{diffV}. 
We only find similar trends as with the GAMA data for the volume V2. The strongest discrepancy is observed for the volume V3, which shows an enhancement in SFR, and higher gas metallicity with respect to the control sample. It is worth noting however, that V3 is also the volume with the smallest number of galaxies in both GAMA and IllustrisTNG. Furthermore, the sample of massive and low metallicity galaxies in the control sample of V3 strongly biases the fit to the zero point, and hence the metallicity difference we measure here is not reliable.\\

Illustris and IllustrisTNG have successfully reproduced the observational trends for individual or small samples of groups and clusters \citep[][respectively]{Genel16,Vogelsberger19}, and some authors have found evolutionary effects, for different redshifts, on the M-Z \citep[][]{Torrey2019} and M-SFR \citep[][]{Torrey2018, Hwang2019} relations. However, these studies explore overall values, and not \mbox{differences}, as we do, over a wider range of redshift, and hence a comparison with our results is not directly applicable.

To explain the discrepancies between IllustrisTNG and GAMA, we list the differences between both samples.
IllustrisTNG provides a redshift (or time) discrete sample, whereas the observations cover a continuous range of values. Moreover, in the IllustrisTNG simulation we observe (or follow) the same galaxies for all the redshift samples. This is obviously not the case in the GAMA observations. Additionally, the selection of star-forming galaxies using the BPT diagram is only possible for GAMA. While for the IllustrisTNG simulations we selected only star forming particles and excluded passive galaxies. Finally, \citet[][]{Zhao2020} found differences in the SFR for the different resolution levels of TNG100 and TNG300.

\begin{table}
\centering
\caption{Redshift, luminosity distance and absolute magnitude limits (r--band, according to the GAMA apparent limiting magnitude m$_r$ = 19.8) for the different simulation snapshots}
\begin{tabular}{lrrrrr}
\hline
\hline
Vol$_{\rm G}$ & z$_{\rm G}$ median & z$_{\rm TNG}$ & Snapshot & d$_{\rm lum}$ [Mpc] & M$_{\rm r}$ \\ \hline
V1 & 0.0854 & 0.08 & 92 & 358.1 & -17.97 \\
V2 & 0.1836 & 0.18 & 85 & 859.4 & -19.87 \\
V3 & 0.2925 & 0.30 & 78 & 1530.8 & -21.12 \\
\hline
\end{tabular}
\label{tab:Vsimul}
\end{table}

\begin{table}
\centering
\caption{Stellar mass statistic for the volume limited samples}
\begin{tabular}{llrrrrr}
\hline
\hline
Vol & & \multicolumn{5}{c}{Mass [log M$_{\odot}$]} \\ \cline{3-7} 
 & & \multicolumn{1}{c}{Min} & \multicolumn{1}{c}{Quartile 1} & \multicolumn{1}{c}{Median} & \multicolumn{1}{c}{Quartile 3} & \multicolumn{1}{c}{Max} \\ \cline{1-1} \cline{3-7}
V1 & & 8.495 & 9.853 & 10.088 & 10.342 & 12.035 \\
V2 & & 9.157 & 10.126 & 10.286 & 10.519 & 12.035 \\
V3 & & 9.870 & 10.420 & 10.521 & 10.683 & 11.470 \\
\hline
\end{tabular}
\label{Vsamples_massStat_simul}
\end{table}

\section{Summary and Conclusions}\label{Conclusion}

We present an analysis of variations in SFR, sSFR and gas metallicity for group galaxies in the GAMA survey. Groups are selected using the friends-of-friends algorithm as described in the G$^3$C GAMA groups catalogue \citep{Robotham11}. The galaxies selected present  at least four emission lines (\Ha, \Hb, [NII]$\lambda$6583 and [OIII]$\lambda$5007). After AGN discrimination by means of the BPT diagnostic, we end up with 26174 star forming galaxies. We generated three volume limited samples to control for evolution and mass variation with redshift due to the Malquimist bias. Variations in the main properties are identified through offsets in the zero point between control and group samples in the M-Z, M-SFR, and M-sSFR relations. Our main conclusions are given in the following bullet points:
\begin{itemize}

    \item The gas metallicity of low redshift galaxies (V1) and low stellar masses is higher than the control sample by $\sim0.05$ dex, 
    while
    group \mbox{galaxies} in the volume V3 show a small decrement in metallicities.
    Our group sample  shows as well a higher metallicity in the highest surface density bin with respect to the control sample by $\sim$0.08 dex, having the rest of the density bins values comparable with those in the field.
    These results are in agreement with \citet{Ellison09}. Stripping of low-metallicity gas from the galaxy outskirts, as well as suppression of metal-poor inflows towards the galaxy centre, or even inflow of pre-processed gas from a rich ICM are key drivers of the enhancement of gas metallicity \citep[e.g.,][]{Bahe2017}.
 \item SFR and sSFR are higher in groups compared to field galaxies in the samples V1 and V2, indicating that mechanisms enhancing SFR are already dominant here, while those quenching them take longer times to be noticeable. 

\item Our full sample was analyzed in bins of group-centric distance, local density, $\Sigma_5$, and stellar mass. We find that the highest enhancements in SFR and sSFR, $\sim 0.3$ and $0.4$, respectively, are found for galaxies with the lower local densities.


\item Contrary to previous authors that find an SFR quenched for galaxies at highest local densities or close group-centric distances, we find small enhacements in SFR of $\sim 0.1$ dex. This difference can be explained as our sample of group galaxies is composed of galaxies with strong emission lines and hence late-type morphologies, excluding the low-SFR and passive systems that drive previous results showing the suppression in SFR.

\item The only signs of quenching in our samples are found for massive galaxies, either when the whole sample is used (Fig. \ref{diffmassV123}), with a volume limited sample of luminous/massive galaxies throughout our whole redshift range (Fig. \ref{diffmassV123_RangeV3All}), or when individual volume limited samples are considered (Fig. \ref{diffV} and \ref{diffZSFR}). Therefore, our data suggests the stellar mass is the benchmark to identify quenching in galaxies with strong emission lines.


\item In contrast, lower gas metallicities are found for galaxies at high group-centric distances (Fig. 11). It is likely that this is connected to the accretion of HI rich galaxies residing preferentially in the outskirts, also responsible for the observed enhancement in SFR at the same  group-centric distances.

\item We tried to reproduce our observational results with group galaxies from the IlustrisTNG simulations, and successfully recovered the general trends for M-Z and M-SFR. However, even though we used the same methodology, we did not find the same quantitative differences as with the observational GAMA data. This discrepancy can be attributed to the discrete outputs for different redshift values in the simulation.

\end{itemize}

\section*{Acknowledgments}

GAMA is a joint European-Australasian project based around a spectroscopic campaign using the Anglo-Australian Telescope. The GAMA input catalogue is based on data taken from the Sloan Digital Sky Survey and the UKIRT Infrared Deep Sky Survey. Complementary imaging of the GAMA regions is being obtained by a number of independent survey programs including GALEX MIS, VST KIDS, VISTA VIKING, WISE, Herschel-ATLAS, GMRT and ASKAP providing UV to radio coverage. GAMA is funded by the STFC (UK), the ARC (Australia), the AAO, and the participating institutions. The GAMA website is http://www.gama-survey.org/. This work was supported by the project ''Evolution
of Galaxies'', of reference AYA2017--88007--C3--2--P, within the  ''Plan
Estatal de Investigaci\'on Cient\'ifica y T\'ecnica y de Innovaci\'on (2017-2020)" of
the "Agencia Estatal de Investigaci\'on del Ministerio de Ciencia, Innovaci\'on y
Universidades". APG is also supported by the Spanish State Research Agency grant MDM-2017-
0737 (Unidad de Excelencia Mar\'ia de Maeztu CAB). 
M.A.L.L acknowledges support from the Carlsberg Foundation via a Semper Ardens grant (CF15-0384).


\section*{Data Availability}

All observational data from the GAMA Project used in this work is publicly available at \url{http://www.gama-survey.org/dr3/}. 
Results from the IllustrisTNG simulation are publicly available at \url{https://www.tng-project.org}.



\bibliographystyle{mnras}
\bibliography{GAMA_groups} 






\bsp	
\label{lastpage}
\end{document}